\documentclass[conference]{IEEEtran}

\usepackage{amsfonts}
\usepackage{textcomp}
\usepackage{xcolor}
\usepackage{fancyhdr}
\usepackage{cite}
\usepackage{graphicx}
\usepackage{psfrag}
\usepackage{subfigure}
\usepackage{url}
\usepackage{stfloats}
\usepackage{amsmath}
\usepackage{array}
\usepackage{fancyhdr}
\usepackage{epsfig}
\usepackage{amssymb}
\usepackage{color}
\usepackage{float}
\usepackage{algorithm}
\usepackage{algpseudocode}
\usepackage{balance}
\usepackage{soul}
\usepackage{caption}
\def\BibTeX{{\rm B\kern-.05em{\sc i\kern-.025em b}\kern-.08em
    T\kern-.1667em\lower.7ex\hbox{E}\kern-.125emX}}
\begin{document}    
\title{An Optimal Unequal Error Protection LDPC Coded Recording System}
\author{\IEEEauthorblockN{Hong-fu Chou}
\IEEEauthorblockA{\textit{Interdisciplinary Centre for Security, Reliability and Trust (SnT), University of Luxembourg, Luxembourg}}
}

\maketitle    

\begin{abstract}
For efficient modulation and error control coding, the deliberate flipping approach imposes the run-length-limited(RLL) constraint by bit error before recording. From the read side, a high coding rate limits the correcting capability of RLL bit error. In this paper, we study the low-density parity-check (LDPC) coding for RLL constrained recording system based on the Unequal Error Protection (UEP) coding scheme design. The UEP capability of irregular LDPC codes is used for recovering flipped bits. We provide an allocation technique to limit the occurrence of flipped bits on the bit with robust correction capability. In addition, we consider the signal labeling design to decrease the number of nearest neighbors to enhance the robust bit. We also apply the density evolution technique to the proposed system for evaluating the code performances. In addition, we utilize the EXIT characteristic to reveal the decoding behavior of the recommended code distribution. Finally, the optimization approach for the best distribution is proven by differential evolution for the proposed system.
\end{abstract}
\section{Introduction}
In recording systems, data bits are transformed into symbol
sequences which must satisfy a run-length-limited (RLL)
constraint.  A $(d,k)$ RLL sequence
\cite{Ayis}\cite{Bergmans}\cite{Robert}\cite{Marcus}\cite{Jin}\cite{Adriaan}
is a sequence for which the number of zero symbols between two
consecutive non-zero symbols is at least $d$ and at most $k$ in
order to alleviate the inter-symbol interference (ISI), and to
assist with timing recovery at the read side
\cite{Moon}\cite{ImminkHollmann}.  Various digital recording systems have applied RLL constraints for the purposes of timing recovery and alleviating inter-symbol interference (ISI). Over optical research in recent years, photochromic materials in \cite{HuPan}\cite{HuXu}\cite{Hu}\cite{Hu2007} have the characteristic of the absorption altering nonlinearly with the exposure energy is applicable for non-saturation
multi-level recording and also allow four-level marks to be written on the disc. Multi-level DVD-ROM driver with conventional setup is studied in \cite{Song} and \cite{Shen}, which raises the intuition of studying over 4-level RLL modulation on DVD-ROM discs in \cite{Hua}. However, it is critical to maintain higher quality of 4-level recording systems, which has raised great concern in \cite{FanCioffi} \cite{Kurkoski} \cite{Vasic} \cite{Lu} \cite{Li_Kumar2005} \cite{Li_Kumar} \cite{Chen_Lin} for combining RLL and error correcting code (ECC).

In order to increase storage reliability, error-correcting codes
(ECC) are usually employed in recording systems. A low-density
parity-check (LDPC) \cite{Gallager}\cite{MacKay} coded recording
system with RLL constraints can be constructed by sequentially
encoding the user data through both the LDPC encoder and the RLL
encoder \cite{ImminkIT}\cite{Hara}. At the read side, an equalizer
followed by an RLL decoder and an LDPC decoder is applied. This is
called the ECC-RLL scheme. However, this arrangement significantly
increases the complexity in cases where turbo equalization
\cite{Douillard} is needed, and is due to the insertion of the RLL
decoder between the equalizer and the LDPC decoder.  Various
approaches have been proposed to overcome this problem
\cite{Vasic}\cite{Li_Kumar2005}\cite{Li_Kumar}\cite{Lu}\cite{Chen_Lin}\cite{Fan}.
In both \cite{Lu} and \cite{Fan}, a reverse-concatenation scheme (or
RLL-ECC scheme) is proposed to enable the construction of RLL codes
with error-correcting capabilities, where the user data is first
encoded through the RLL encoder and then the LDPC encoder. A major
disadvantage of this approach is the rate loss resultant from the
RLL code, although some novel arrangements can be used to reduce
this loss. In \cite{Vasic}, the authors propose to obtain a sequence
that satisfies the RLL constraint by deliberately flipping those
bits which violate the constraint in each LDPC codeword. In the
flipping-based scheme, turbo equalization can still be performed
since there is no RLL decoder at the read side. The scheme relies
only on the capability of the LDPC code to correct the errors caused
by the flipping operation. However, such a scheme can only be
applied to systems with a very loose RLL constraint, otherwise the
number of flipping bits becomes too large, which significantly
degrades the error performance. In order to reduce the number of
flipped bits, a selective flipping technique
\cite{Li_Kumar2005}\cite{Li_Kumar} is employed. The idea is that for
each LDPC codeword, we obtain multiple bit-flipping versions (i.e.,
multiple candidates) and then choose the candidate that generates
the fewest flipping errors when writing. This technique is efficient
except for the need for side information, which is required in order
to inform the read side which candidate has been chosen at the write
side. The side information should be well-protected and should also
satisfy the RLL constraint. In \cite{Chen_Lin}, the RLL constraint
is applied to the output of the LDPC decoder in order to detect the
flipped bits. Then, in the next iteration, the extrinsic LLRs (log-likelihood ratios) are to be passed to the equalizer, and the
LDPC decoders, respectively, are modified by reversing the polarity
of the extrinsic LLR of each of the detected flipped bits. The
detection process presented in \cite{Chen_Lin} is not appropriate
for flipping-based RLL recording systems that are not based on the
selective flipping technique since the output of the LDPC decoder
may contain too many errors, which will make the detection of the
flipped bits extremely difficult.

In this thesis, we focus on an LDPC-coded recording system with the
$(0,k)$ constraint. We propose to modify the method presented in
\cite{Chen_Lin} using two approaches based on the detection of RLL flipped bit. For the first approach, the soft
information of the detected flipped bits is adjusted to alleviate
the effect of hard errors due to the incorrect estimation of flipped
bits. The methods in consideration include (i) erasing the soft
information; (ii) using the average of the absolute values of the
LLRs generated during each iteration as the magnitude of the soft information; and
(iii) using the average of the absolute values of the LLRs generated during each
iteration as the clipping threshold of the soft information. For the
second approach, we consider a design for flipped-bit detection
that properly utilizes both the RLL constraint and the parity-check
constraint of the LDPC code in order to increase the likelihood of the
correct detection of the flipped bits. We show that by appropriately
examining both the RLL constraint and the parity check constraint at
the output of the LDPC decoder for each iteration, we can
significantly increase the probability of a correct detection of the
flipped bits and, hence, reduce the overall error rate of the
system. These two approaches can be combined to further improve the
error performance.

Multilevel (ML) recording techniques achieve high recording densities and data rates in optical data storage systems. The user data stored on the optical disc is a series of ML symbols. Modulation codes transform user data bits into an ML symbol sequence conforming to certain channel constraints to satisfy various purposes.  Over optical research in recent years, photochromic materials in \cite{HuPan}\cite{HuXu}\cite{Hu}\cite{Hu2007} has the characteristic of the absorption altering nonlinearly with the exposure energy is applicable for non-saturation
multi-level recording and also allow four-level marks to be written on the disc.
In optical recording systems, a uniform amplitude segment, or pulse, of the RLL
waveform is represented as a single mark on the recorded optical storage medium.
This mode of recording is known as pulse length modulation (PLM) recording since
the lengths of recorded marks vary in sympathy with the lengths of the pulses that
comprise the RLL waveform. PLM recording requires that individual marks be written with well-controlled length and furthermore, that the mark's lengths be reliably
discriminated when the recorded medium is subsequently read. A PLM sequence encoded from the RLL encoder has pulses with a minimum
length corresponding to the shortest run ($d + 1$ channel digits) and a maximum length corresponding to the longest run ($k +1$ channel digits). The pulses are one of $M$ amplitude levels which are obtained by adding a cyclic (modulo-$M$) , for binary $M = 2$,
incremental amplitude level between the present symbol and last pulse. This limits
the propagation of errors in recovered pulse amplitudes since the relative amplitude
values of only two adjacent pulses are required to determine the amplitude of the
most recently recovered pulse. It also provides some immunity to large spatial-scale
reflectance variations that could be caused by fingerprints on the storage medium
protective surface, birefringence in the medium substrate and focus/tracking perturbations that may occur while the storage medium is being read. A Non-Return-
to-Zero Inverted (NRZI) symbol sequence $z_{i}$ is obtained by mapping a RLL symbol
sequence $x_{i}$ over the PLM precoder. The mapping is defined as
\begin{equation}
z_{i}=x_{i}\bigoplus z_{i-1}
\end{equation}
The symbol sequence $z_{i}$ after PAM signal mapping will be recorded on the disc by
controlling the power of the writing laser.
Multi-level DVD-ROM driver with conventional setup is studied in \cite{Song} and \cite{Shen}, which raise the intuition of studying over 4-level RLL modulation on DVD-ROM discs in \cite{Hua}. However, it is critical to maintain higher quality of 4-level recording systems, which has raised great concern in \cite{Chen_Lin} \cite{FanCioffi} \cite{Kurkoski} \cite{Vasic} \cite{Lu} \cite{Li_Kumar2005} \cite{Li_Kumar} for combining RLL and error correcting code (ECC).
Following the thread of investigating LDPC code in \cite{Kasai}, \cite{Poulliat}, and \cite{Hossein}, we achieve unequal error protection (UEP) properties by designing the variable and checking node degree distribution of the code in an irregular approach. The connection degree of the variable nodes leads to different types of error-correcting ability. The codeword bits are divided into several parts according to the connection degree and each class has a different protection capability. It is known that it is best
to have high degrees for variable nodes. The more information a variable node receives from its adjacent check nodes, the more accuracy it judges about its correct value. For the finite-length UEP LDPC code in \cite{Nazanin}\cite{Nazanin2}, the authors optimized the degree distributions by unequal density evolution formulas over the binary erasure channel. They presented the merit of the proposed design methodology for unequal protection and provided a recommended node distribution with a rate 1/2 under performance guarantee.  In \cite{Neele}, they investigated the optimization for both UEP LDPC code and signal labeling. They proposed a flexible code design algorithm for several protection classes and arbitrary modulation schemes using Gray mapping. It revealed that appropriate code reduces the overall bit-error rate. The optimization process of the degree distribution for higher-order constellations is provided by optimizing each protection class after another by linear programming. Thus, the optimization target is to find a variable node degree
distribution for the whole code that maximizes the average variable node degree of the class being optimized.
Owing to the well-designed coded modulation for unequal error protection, we can design the UEP LDPC code to cooperate with modulation and retrieve the flipped error during iterative decoding.

In this thesis, we proposed a decoding scheme using UEP LDPC code by means of regular interleaver to confine the occurrence of flipping error to a section of the codeword. The specified section with flipping error is attributed to a high-degree node with high error protection capability. Signal labeling is designed to enhance the Euclidean separation for robust unequal protection. Iterative decoding within several inner and outer iterations can solve the hard error on the read side. Density evolution and EXIT characteristics are applied to provide the intensive analysis. The optimization for node distribution using differential evolution is investigated for the proposed scheme.

The thesis is organized as follows. In section II,  a brief revisit of the flipping-based LDPC-coded system proposed in \cite{Vasic}.  In section III, unequal protection LDPC code is introduced in an irregular LDPC code approach. We depicted two types of 4-level RLL recording schemes using deliberate flipping. Error floor is alleviated by the proposed scheme using appropriate node distribution. In section IV, density evolution and EXIT characteristic are applied to analyze the recording scheme, and simulation results are presented to illustrate the advantage of recommended node distribution. The optimization approach using differential evolution searches for the best node distribution for the proposed scheme. Finally, Conclusions are given in section V.

\section{An RLL-constrained LDPC Coded Recording System Using Deliberate Flipping }
The idea of deliberately flipping some bits of the
LDPC codeword to meet the RLL constraint at the write side, and
using the error-correcting capability of the LDPC code to remove the
flipped bits at the read side, was proposed in \cite{Vasic}. Improved versions
were presented in \cite{Li_Kumar} and \cite{Chen_Lin}.
\begin{figure}[ht]
\centering
\includegraphics[width=0.5\textwidth]{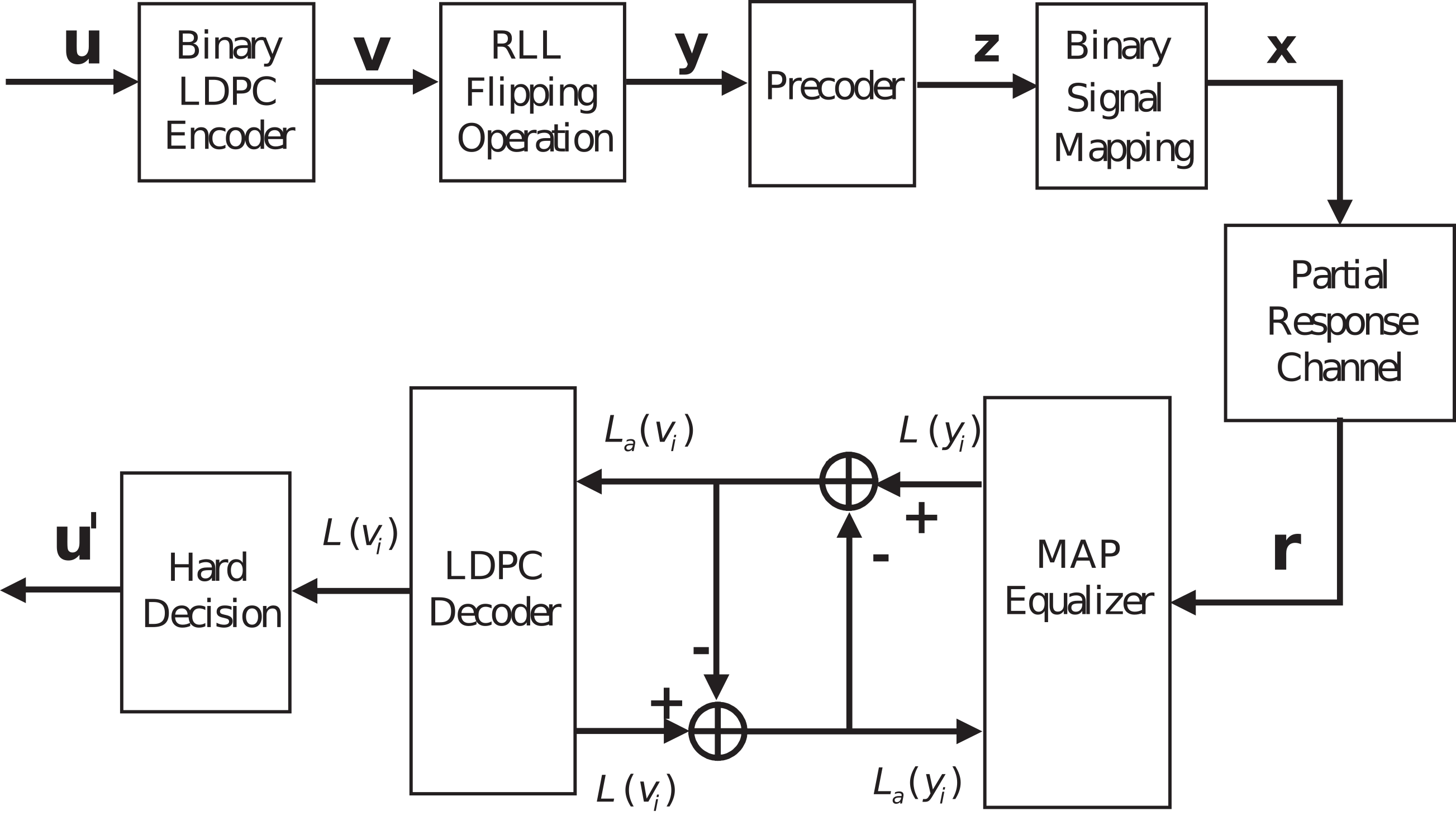}
\caption{An LDPC-coded binary recording system based on RLL constraints using deliberate flipping.}
\label{btx}
\end{figure}
\subsection{Deliberate Flipping at the Write Side}
Fig.~\ref{btx} shows the block diagram for the system presented in
\cite{Vasic}. A $K$-bit message $\textbf{u}$ is forwarded to the
encoder of a binary ($N$, $K$) LDPC code to produce an $N$-bit
codeword $\textbf{v}\equiv$($v_1$,$v_2$,$\cdots$,$v_{N}$). The
codeword $\textbf{v}$ is forced to satisfy the $(0,k)$ RLL
constraint by flipping certain bits of $\textbf{v}$ that violate the
$(0,k)$ constraint, thereby obtaining a flipped version denoted as
$\textbf{y}\equiv$($y_1$,$y_2$,$\cdots$,$y_{N}$), where
$v_i\in\{0,1\}$, $y_i\in\{0,1\}$, $i=1,2,\cdots,N$.  The flipping
operation can be implemented as the summation of the codeword
$\textbf{v}$ and a binary location vector
$\textbf{q}$$\equiv$($q_1$,$q_2$,$\cdots$,$q_{N}$), i.e.,
$\textbf{y}=\textbf{v}\bigoplus \textbf{q}$, where $\oplus$
represents the XOR operation. The location vector can be obtained
using a sliding-window-based algorithm. For the case of the $(0,k)$
constraint considered in this work, the steps are described as
follows:

\begin{description}
\item[Step 1:]\hspace{0.1cm} The window size is set to $k+1$ bits.
                            The $N$-bit location vector $\textbf{q}$
                            is set to the zero vector, and $i=1$.
\item[Step 2:]\hspace{0.1cm} If both vectors $(v_i,v_{i+1},\cdots,v_{i+k})$
                            and $(q_i,q_{i+1},\cdots,q_{i+k})$ are the zero $(k+1)$-tuple,
                            then set $q_{i+k}$ to $1$.
\item[Step 3:]\hspace{0.1cm} Repeat Step 2 for $i=2,3\cdots,N-k$.
\end{description}

If $q_i=1$, then $v_i$ will be flipped before being stored
in the recording media.

In Fig.~\ref{btx}, the output of the RLL flipping operation unit,
$\textbf{y}$, is processed by a precoder to yield $\textbf{z}$ =
$(z_1, z_2, \cdots, z_N)$, where
\[
z_i = y_i + z_{i-1} \mbox{ (mod 2)}.
\]
Through a modulator (binary signal mapper), we have $\textbf{x}$ =
$(x_1, x_2, \cdots, x_N)$, where
\[
x_i = (-1)^{z_{i}}.
\]

\subsection{Channel Model for Magneto-Optical Recording Channel}
The signal $\textbf{x}$ is written to the recording media.
First, we introduce a
practical magneto-optical (MO) binary recording channel presented in
\cite{Hongwei}\cite{hans1} is considered in this study.
The channel impulse response $f(t)$ considered in \cite{Hongwei} is given by
\begin{equation}
\label{15} f(t) = \frac{2}{ST\sqrt{\pi}} exp\{-(\frac{2t}{ST})^2\}
\end{equation}
where $T$ is the bit duration and $S=1.0$ is used for binary channel. If we let
$\prod(t)$ be the function which equals 1 for $0 \le t \le T$ and
equals 0 otherwise, then the transition response $g(t)$ is the
derivative $f(t)$ as to be the impulse response of the channel. The output signal of the MO
recording channel is
\begin{equation}
\label{16} r(t) = \sum_{i=-\infty}^{\infty}
x_{i}g(t-(i+\Delta_{i})T) + n(t),
\end{equation}
where $\{\Delta_{i}\}$ is an independent and identically
distributed random process representing the jitter effect, and
$n(t)$ is the additive white Gaussian  noise (AWGN) with one-sided
power spectral density $N_{0}$.  The jitter noise is an important noise source of optical storage systems. It is
induced by transition jitter, which causes written transitions to shift with respect
to their nominal positions. The jitter value $\{\Delta_{i}\}$ is defined as the root mean square (RMS) transition shift normalized in channel bit period $T$. We assume that each $\Delta_{i}$
is a zero mean Gaussian random variable with variance
$\sigma^{2}_{\Delta}$.  By assuming that $|\sigma_{\Delta}|T$ is
small and $g(t)$ is sufficiently bandlimited \cite{HsinYi2010}, we
have
\begin{equation}
\label{17} g(t-(i+\Delta_{i})T) \approx g(t-iT) -
\Delta_{i}Tg'(t-iT).
\end{equation}
By substitution, (\ref{16}) then becomes
\begin{eqnarray}
\label{18} r(t) &\approx& \sum_{i=-\infty}^{\infty} x_{i}g(t-iT) -
\sum_{i=-\infty}^{\infty} x_{i}\Delta_{i}Tg'(t-iT)+n(t)
                                            \nonumber\\
                &\approx& \sum_{i=-\infty}^{\infty} x_{i}g(t-iT) -
\sum_{i=-\infty}^{\infty} x_{i}\Delta_{i}Tf(t-iT)+n(t)
                                            \nonumber\\
                &=&x(t)+z(t)+n(t),
\end{eqnarray}
where intersymbol interference (ISI) term is $x(t)=\sum_{i=-\infty}^{\infty}x_{i}g(t-iT)$, the
jitter noise is $z(t)=\sum_{i=-\infty}^{\infty}x_{i}\Delta_{i}Tf(t-iT)$ and white Gaussian noise $n(t)$. The average energy of the jitter noise for each received
code symbol is
\begin{eqnarray}
\label{19} M_{0} &=&
E\{x_{i}^2\}E\{\Delta_{i}^2\}E\{\sum_{j=\infty}^{\infty}[Tf(jT)]^2\}
                           \nonumber\\
                 &=&
                 \sigma_{\Delta}^{2}\sum_{j=\infty}^{\infty}[Tf(jT)]^2.
\end{eqnarray}
The jitter noise effect is simulated according to (\ref{18}) using a
Gaussian distributed $\Delta_{i}$.  In the simulation, the energy
of the signal $x_{i}$ is fixed as 1, and the energy of the jitter noise
$M_{0}$ is set to be proportional to $N_0$. We set $M_0$ = $\beta
N_0$, where $\beta=0.15$. In the simulation, discrete-time samples are
required. From $f(t)$ and $g(t)$, we have
$Tf(-2T)$ = $Tf(2T)$ = 0.0962, $Tf(-T)$ = $Tf(T)$ = 0.1085, $Tf(0)$ = 0.1128
and $g(T)$ = $g(4T)$ = 0.1114, $g(2T)$ = $g(3T)$ = 0.1028. In
addition, we set $f(iT)$ = 0, for $\mid i \mid
> 2$ and $g(iT)$ = 0 for $i \le 0$ and $i \ge 5$.  Using (\ref{18}),
we can obtain the associated discrete-time output sequence denoted as $\textbf{r}$.

For the multilevel recording channel discussed in \cite{Hua} and \cite{HsinYi2010}, they investigated that the PR model is closed to the actual ML-RLL DVD channel with transfer function (\ref{15})  of the $S$ = 4.6 channel. For theoretical experiments, the PR channel has impulse response $\overline{h}$=(1,2,2,1) denoted as PR(1,2,2,1).
 From $f(t)$ and $g(t)$ of 4-level recording channel with jitter noise, we have $Tf(-2T)$ =$Tf(2T)$ =
0.0981, $Tf(-T)$ = $Tf(T)$ = 0.2251, $Tf(0)$ = 0.2969.  We also
have $g(T)$ = $g(4T)$ = 0.1601, $g(2T)$ = $g(3T)$ = 0.2727. For
simplicity, we set $f(iT)$ = 0, for $\mid i \mid > 2$ and $g(iT)$ = 0
for $i \le 0$ and $i \ge 5$. In the simulation, a MAP
detector with $4^3$ states will be used for this optical recording
channel, where the state of the precoder is integrated into the state
of the channel represented by $g(iT)$.
\subsection{Turbo Equalization at the Read Side}
The sequence $\textbf{r}$ suffers not only from noise (AWGN and
jitter noise) but also from intersymbol interference (ISI). As
shown in Fig.~\ref{btx}, the sequence $\textbf{r}$ is forwarded to a
SISO (soft-input soft-output) MAP (maximum $a$ $posteriori$)
equalizer combined with the precoder and a PR (partial response)
channel in order to combat the effect of ISI. The MAP equalizer
utilizes the BCJR algorithm \cite{MAP} to produce the soft outputs
(log-likelihood ratios, LLRs) $L(y_i)$ for bits $y_i$, where
$i=1,2,\cdots,N$. The $a$ $priori$ LLRs $L_a(v_i)$, where
$i=1,2,\cdots,N$, for the LDPC decoder are calculated according to
$L_a(v_i)=L(y_i)-L_a(y_i)$, where $L_a(y_i)$ is the $a$ $priori$ LLR
for the MAP equalizer and is produced from the soft output $L(v_i)$
of the LDPC decoder according to $L_a(y_i)=L(v_i)-L_a(v_i)$. The
sum-product algorithm presented in \cite{MacKay} using $U_{i}$ inner
iterations is performed in the LDPC decoder. Moreover, $U_{o}$ outer
iterations are processed between the MAP equalizer and the LDPC
decoder in order to realize the turbo equalization \cite{Douillard}.

\section{Unequal Error Protection Technique for RLL-constraint LDPC Coded Recording System }
In certain communication systems, the coding scheme is designed in such a
way that the most important information bits have a lower error
rate than other information bits. Codes that are designed to provide different levels of data protection are known as unequal error protection (UEP) codes. UEP codes were first studied by Masnick and Wolf \cite{wolf} and later by others.  Each coded signal sequence in correspondence to a most important message part is associated with a cloud. The mapping of information bits to
coded signal sequences is made in such a way that the minimum
distance between coded signal sequences in different clouds
is larger than the minimum distance between coded signal sequences
within a cloud. This is an unequal error protection
(UEP) coding scheme. While such a coding scheme is used, different code levels provide different degrees of protection for different parts of a transmitted information sequence. It is necessary to
combine coding and modulation in such a way that the required graceful degradation in BER performance is achieved through error control coding. This approach is used to solve the RLL flipping error at the read side.

The multi-level coded modulation for UEP system guidelines is presented in \cite{Morelos}\cite{IsakaLinFoss1}. A $2^{M}$-ary modulation signal set $S$ is partitioned into $M$ levels.
At the $i$-th level, for $1\leq i\leq M$, the signal set is divided into two subsets $S_{i}(0)$ and $S_{i}(1)$, such that the intraset square Euclidean distance (SED) $\delta_{i}^{2}$ is maximized. The binary code $C_{i}$ for $1\leq i\leq M$ is selected to satisfy the
the following inequality,
\begin{equation}
d_{1}\delta_{1}^{2}\geq d_{2}\delta_{2}^{2}\geq ...\geq d_{M}\delta_{M}^{2}
\end{equation}
where $d_{i}$ is the minimum Hamming distance of linear block code $C_{i}$ for $1\leq i\leq M$.
The Euclidean separations of the $i$-th level is defined as $d_{i}\delta_{i}^{2}$. The above principle is useful in specifying the asymptotic
error performance of a UEP-coded modulation. Also, the partitioning process leads to the labeling of the signal sets. In Fig.\ref{partitioning}, Ungerboeck partitioning of signal points
are labeled with strings of symbols from a certain finite alphabet,
mostly the binary alphabet {$0,1$}.
The conventional signal set partitioning for maximizing intraset distance at each level is ideal for constructing modulation codes for one-level error protection. However, multilevel codes using Ungerboeck partitioning produce a large increase in effective error coefficients in the first several decoding stages and do not perform well in a UEP system. Additionally, the block partitioning can be carried out independently and simultaneously in parallel which removes the error propagation from the first decoding stage to the second decoding stage.
Hence, the upper bound of error probability using block partitioning of the first and second decoding stages allows the error probability to be written as \cite{FossorierRhee},
\begin{equation}
P_{e}(b_{i})\leq \sum_{w=d_{i}}^{n}w/n A_{w}2^{-w}\sum_{i=0}^{w}{n\choose w}Q(\sqrt{2RE_{b}/N_{0}d^{2}_{P}(i)})
\end{equation}
where $w$ denotes the Hamming weight of an incorrectly decoded codeword, $d^{2}_{P}(i)$
is the smallest SED from the corresponding point $P$ to the error event and $A_{w}$ is the number of the codeword with weight $w$. In order to maintain the UEP capability, the strategy is to reduce
the number of nearest neighbors ($NN$) at every partition level. As
a result, only a small number of signal points, those located
near the decision boundary, will have neighbors at the minimum
distance. Particularly, conventional Ungerboeck symmetrical partitioning maximizes intraset
SED ($\delta_{i}^{2}$) at each level, which does not perform well for the UEP system. Since the
number of $NN$ associated with the $i$-th level using block partitioning is much less
than when using Ungerboeck partitioning, this results in the upper bound of error probability:
\begin{equation}
P_{e}(b_{i})\leq \sum_{w=d_{i}}^{n}w/n A_{w}2^{w}Q(\sqrt{2RE_{b}/N_{0}w\Delta_{1}^{2}})
\end{equation}
where $\Delta_{1}^{2}$ is the smallest SED of the first level.
 Owing to the large increase in effective error coefficients in the first signal level $2^{w}$, this causes the UEP inequality (4.1) to be destroyed and also causes system performance degradation in equation (4.3). Furthermore, the increase in the error coefficient has a direct relationship to the poor performance such that Ungerboeck partitioning is not appropriate for UEP systems.
We organize the properties of a well-designed UEP-coded system as follows:
\begin{figure*}[hbt]
\begin{center}
\includegraphics[width=0.8\textwidth]{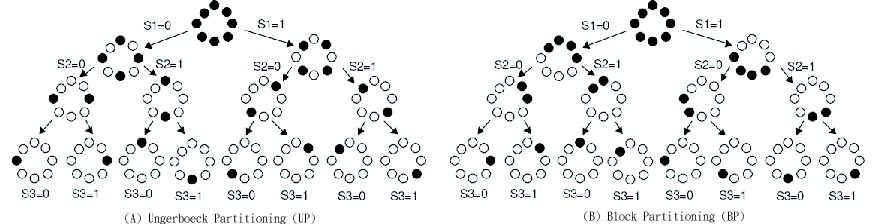}
\caption{Partitioning strategies for an 8PSK constellation.}
\label{partitioning}
\end{center}
\end{figure*}

\begin{description}
\item[1)] \hspace{0.5 cm}
The unequal error protection for various parts of a message must satisfy the inequality (4.1). This condition ensures that two messages are identical in the first $i-1$, but different in the $i$-th part. Then, their corresponding codewords are separated by a squared Euclidean distance of at least $d_{i}\delta_{i}^{2}$, which is called the Euclidean separation.
\item[2)] \hspace{0.5 cm}
The signal constellation possesses multi-level error protection, meaning that
maximizing intraset SED is no longer valid for UEP system design. The design principle must decrease the number of $NN$s at every partition level in order to enhance the robust protection part of the error-correcting code. This strategy leads to a larger difference in Euclidean separation between each level, which implies that more Euclidean separation on each level provides better performance for UEP systems.
\item[3)] \hspace{0.5 cm}
The error coefficient dominates the small-to-medium SNRs. The performance of a multi-level system using Ungerboeck partitioning is determined by the error coefficient $2^{d}\cdot A_{d}$, where $d$ denotes the minimum Hamming distance, and $A_{d}$ is the number of codewords with Hamming weight $d$. The large increase in error coefficient degrades the error performance exponentially growth. However, if we partition the signal to decrease the number of $NN$s, as the block partitioning, the error coefficient becomes $2^{-d}\cdot A_{d}$. The loss of intraset SED is compensated for by the drastic reduction in error coefficients $2^{2d}$.
\end{description}

Now, the irregular LDPC codes with unequal error protection are introduced.
Let an $M\times N$ matrix, denoted as $\mathbf{H}$, be a
parity-check matrix of an ($N$, $K$) LDPC code, where $M\geq (N-K)$.
The parity-check matrix $\mathbf{H}$ can be represented by a Tanner
graph \cite{Tanner}, which is a bipartite graph consisting of
variable (bit) nodes and check nodes representing the columns and
rows of $\mathbf{H}$, respectively. If there is a $1$ at both the $i$-th
column and the $j$-th row of $\mathbf{H}$, then there is an edge
between the $i$-th variable node and the $j$-th check node in its
Tanner graph. Hence, parity-check matrix $\mathbf{H}$ has maximum $d_{v}$ and $d_{c}$
$ 1$ in each column and in each row. For a UEP LDPC
code, ($d_{v}$,$d_{c}$)-irregular LDPC codes can form different error
protection capabilities within a block of codewords. According to \cite{Richard}, $\lambda$ and
$\rho$ are denoted as the variable and the check degree distribution respectively, where
\begin{equation}
\lambda(x)=\sum_{k}\lambda_{k}x^{k-1}
\end{equation}
and
\begin{equation}
\rho(x)=\sum_{l}\rho_{k}x^{k-1}
\end{equation}
for
a $k$-th variable node degree ($VND$) or $l$-th check node degree ($CND$).
From the node perspective, $\delta$ and $\gamma$ are denoted as the variable and the check node distribution, respectively,
for which
\begin{equation}
\delta_{k}=\frac{\lambda_{k}/k}{\sum_{k}\lambda_{k}/k}
\end{equation}
and
\begin{equation}
\gamma_{k}=\frac{\rho_{l}/l}{\sum_{l}\rho_{l}/l}.
\end{equation}
We consider two classes of protection capability, one of which is the strong correcting part where $VND=4$,$5$,$6$ or $7$,
and the other is the weak correcting part where $VND=2$ or $3$.

\subsection{The 4-level RLL Coding Scheme for Unequal Protection}
\begin{figure}[hbt]
\begin{center}
\includegraphics[width=0.5\textwidth]{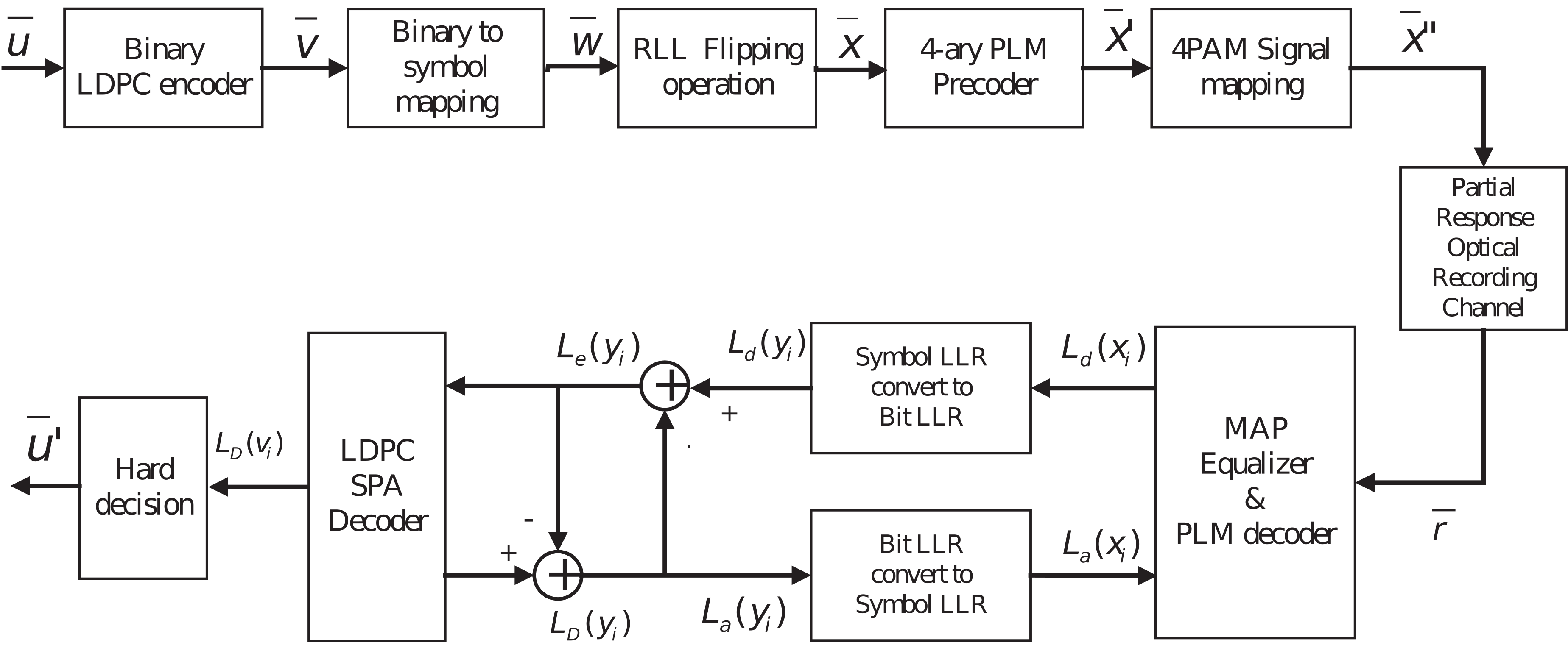}
\caption{4-level LDPC coded recording system with RLL constraints based on deliberate error insertion}
\label{traditionrxtx}
\end{center}
\end{figure}

 Fig.\,\ref{traditionrxtx} shows the system block diagram of a 4-level recording scheme. The message $\overline{u}$=(u$_{1}$,u$_{2}$,...,u$_{m}$) carries m bits for a block as an input to the LDPC encoder. The output, $\overline{v}$=(v$_{1}$,v$_{2}$,...,v$_{N}$), of the encoder is an n-bit LDPC codeword and is necessary in order to translate the binary codewords into symbol format. We use $\overline{w}$=(w$_{1}$,w$_{2}$,...,w$_{\frac{N}{2}}$) to represent the symbol form of codeword $\overline{v}$. However, the sequence $\overline{w}$ can possibly violate the RLL constraints. Hence, the RLL flipping operation translates the symbol sequence $\overline{w}$ to sequence $\overline{x}$. In order to obtain the sequence $\overline{x}$ with RLL constraints, sequence $\overline{w}$ changes each first violated symbol 0 to a nonzero symbol among the run length of consecutive zeros. The sequence $\overline{x}$ is passed into a 4-level PLM precoder and follows the rule of x$'_{i}$=x$'_{i-1}$$\oplus$x$_{i}$ (mod4) to obtain the sequence $\overline{x}'$, where $\oplus$ represents the XOR operation. Before writing data to the media, 4-level signal mapping translates $\overline{x}'$ into $\overline{x}''$, which maps symbols $\{$0,1,2,3$\}$ to the signal level ($-3/\sqrt{5}$, $-1/\sqrt{5}$, $1/\sqrt{5}$, $3/\sqrt{5}$). The PR channel with impulse response $\overline{h}$=(h$_{0}$,h$_{1}$,...,h$_{p-1}$) and AWGN noise deteriorates the sequence $\overline{x}''$ from sequence $\overline{r}$ attained at the reading side. To compensate for the ISI effect, a symbol-based MAP detector (channel equalizer) computes the soft channel value using the BCJR algorithm. Then, the symbol LLR value is transformed into a bit LLR value so as to forward it to the binary LDPC decoder. The LDPC decoder, based on using belief propagation (BP) algorithm \cite{MacKay}, uses a soft channel value to decode the $a$ $posteriori$ LLR (log likelihood ratio) within $U_{i}$ iterations. Soft iterative decoding between the MAP channel equalizer and the LDPC decoder decodes the message sequence $\overline{u}$ from sequence $\overline{r}$. After $U_{o}$ iterations, the recorded information is recovered in order to compare sequence $\overline{u}'$ with the original sequence $\overline{u}$.

\begin{figure}[hbt]
\begin{center}
\includegraphics[width=0.5\textwidth]{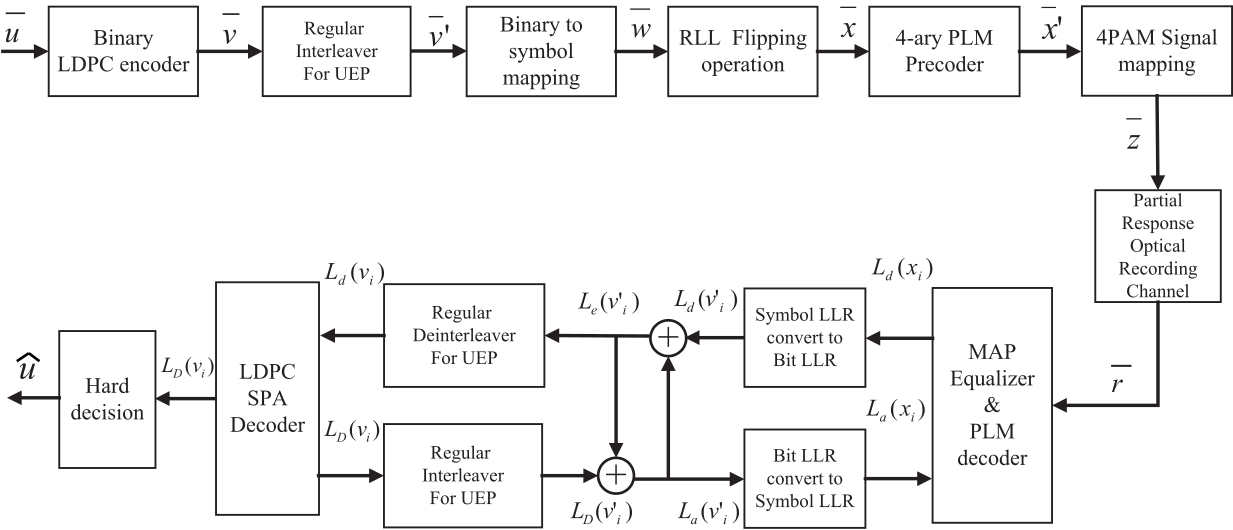}
\caption{4-level LDPC coded recording system with RLL constraints  using UEP LDPC code to recover flipping errors}
\label{ueprxtx}
\end{center}
\end{figure}

For the 4-level recording system, we illustrated the proposed system block diagram in Fig.\,\ref{ueprxtx}. The write side sequentially processes the message data via a binary LDPC encoder, a regular interleaved for unequal protection, binary to symbol mapping, the RLL flipping operation, a 4-level PLM precoder, and 4-level pulse length modulation (PAM) signal mapping. Then, the transmission sequence passes through the partial response channel and the deinterleaver reverses the interleaving manner. The reading side consists of a regular deinterleaver, an interleaver, and an iterative decoding between the MAP channel equalizer and the LDPC decoder. We utilize the non-reset decoder since the error correction capability is more robust than that of the reset decoder.
\label{UEP}

\begin{figure}[hbt]
\begin{center}
\includegraphics[width=0.5\textwidth]{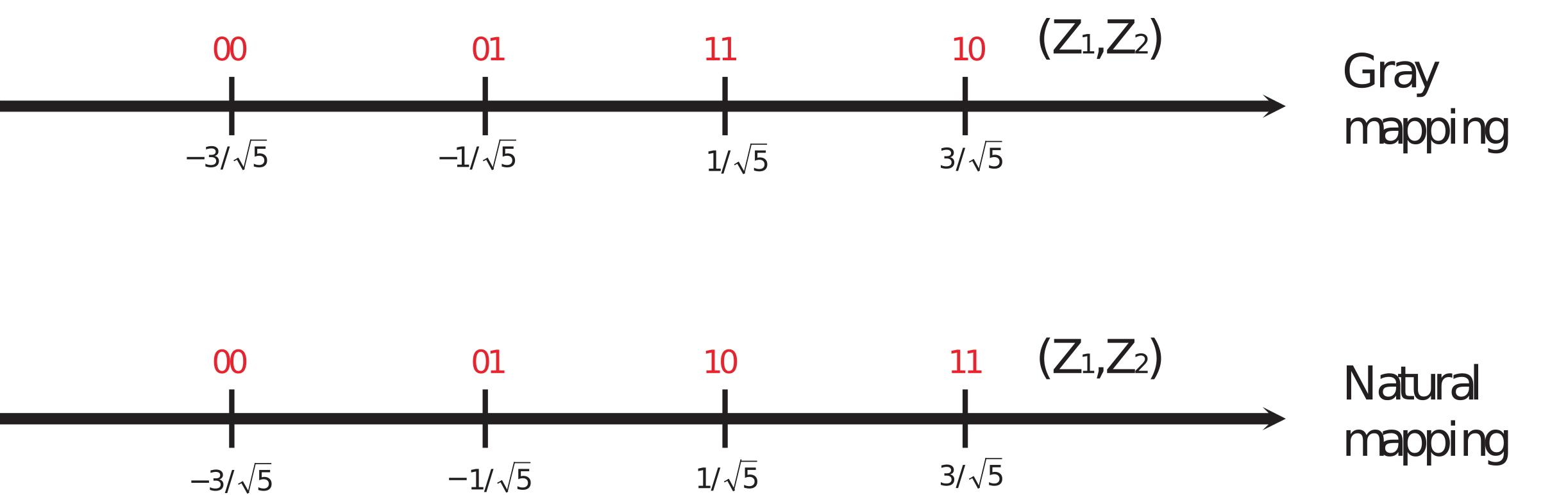}
\caption{Signal labeling for 4-level PAM}
\label{uep1}
\end{center}
\end{figure}

For 4-level PAM, the corresponding two signal labeling bits are denoted as ($Z_1$, $Z_2$). We consider both natural mapping and Gray mapping are considered, as shown in Fig.~\ref{uep1}, along with their associated labels. The average Euclidean weight enumerator (AEWE) is defined as
\begin{equation}
\Delta^{2}_{e}(X)=(1/M)\sum_{z}X^{||f(z)-f(z \oplus e)||^{2}}
\end{equation}
where $e$ is the error vector $e=\overline{r}\oplus\overline{z}$, and there are M pairs of signal vectors $\overline{r}$ and $\overline{z}$ such that $\overline{z}$=$\overline{r}\oplus e$. The AEWE for both natural mapping and Gray mapping is shown in Table~\ref{ueptable1}.

\begin{table}[ht]
\centering
\caption{The AEWE of Gray mapping and natural mapping.}
\begin{tabular}[b]{|c|c|c|}
\hline
e&Gray mapping$\Delta^{2}_{e}(X)$&Natural mapping$\Delta^{2}_{e}(X)$ \\
\hline
$00$&$X^{0}$&$X^{0}$ \\
\hline
$01$&$X^{0.8}$&$X^{0.8}$ \\
\hline
$10$&$\frac{1}{2}X^{0.8}+\frac{1}{2}X^{7.2}$&$X^{3.2}$ \\
\hline
$11$&$X^{3.2}$&$\frac{1}{2}X^{0.8}+\frac{1}{2}X^{7.2}$ \\
\hline
\end{tabular}
\label{ueptable1}
\end{table}
 For the proposed UEP design technique, the write side allows the RLL flipping configuration to be set and allows the freedom of flipping the non-zero symbol $1$ or $2$ corresponding to signal bit ($Z_1$=0, $Z_2$=1) or ($Z_1$=1, $Z_2$=0).  Two different Euclidean distances for the flipping error ($Z_1$=0, $Z_2$=1) or ($Z_1$=1, $Z_2$=0) correspond to long-distance flipping or short-distance flipping respectively.  We observe that the $\Delta^{2}_{10}(X)$ for Gray mapping is $\frac{1}{2}X^{0.8}+\frac{1}{2}X^{7.2}$ and the $\Delta^{2}_{10}(X)$ for natural mapping is $X^{3.2}$ with a distance of 3.2 and $\Delta^{2}_{01}(X)=X^{0.8}$ with a distance of 0.8. Since the long-distance flipping for Gray mapping is equal to 7.2, it seems that it is difficult for the read side to recover this type of RLL flipped bits, despite the fact that Gray mapping has the better nearest neighbor in the signal labeling configuration($NN$($Z_1$)=1 and $NN$($Z_2$)=2). Consequently, Gray mapping leads to a worse long-distance flipping with a distance of 7.2 being set compared to natural mapping with a distance of 3.2. Hence, we selected natural mapping for the flipped bit configuration as the Euclidean distance is equal to 0.8 for short-distance flipping and 3.2 for long-distance flipping. We identify the proposed system using natural mapping that has better flipping error recovery properties in section III E. \\

\subsection{The Proposed Unequal Protection Approach}
For a $2^{M}$-level LDPC coded system, $2^{M}$-level (or $2^{M}$-ary)
sequence with symbol alphabet, $\{0,1, ..., 2^{M}-1\}$, where $1 \le M
< \infty$. Let $Z$=$(Z_1,Z_2,...,Z_M)$ be the bit labeling for a transmitted symbol where, for $1\leq i\leq M$, $Z_i$ is the $i$-th part of the LDPC code bit. The RLL flipping operation is configured such that consecutive zeros are changed to the non-zero symbol corresponding to the flipped $j$-th part of the bit labeling, where $j\in1,2...,M$.
Now, we present the proposed UEP scheme as follows.

\subsection{Proposed UEP Scheme Type I}

For an LDPC $(N,K)$ code, binary sequence $\overline{v}$ is forwarded to a regular interleaver, where the bit ordering of type I is as follows:
\begin{eqnarray}
\pi_{I}(l)=\left\{
            \begin{aligned}
            & \{M\cdot l\}mod~(N-1),~0\leq l\leq N-2  \\
            &  l,~l=N-1
            \end{aligned}
         \right.
\end{eqnarray}
where $\pi_{I}$ denotes the ordering operation of type II regular interleaver. This interleaving allows a segment of code bit on a specific $i$-th part of the bit labeling to be allocated. Since the UEP property possesses a strong correction segment, the $j$-th part of the flipped bit is designed to interleave on the strong part of the code bit. An example of a 4-level UEP recording system block diagram is illustrated in Fig.\,\ref{ueprxtx}.
On the write side, we are able to set the RLL flipped bit on the signal labeling $Z_1$ or $Z_2$. If the signal labeling $Z_2$ is individually set to be flipped, and type I interleaving is applied, the proposed technique regularly interleaves the binary signal $Z_1$ and $Z_2$ respectively to form a half of codeword with the flipping error ($N/2\leq l\leq N-1$) and another half without flipping error ($0\leq l\leq N/2-1$).  Therefore, this approach allows the occurrence of flipped bits to be limited within a codeword segment. For a 4-level system, we design a UEP LDPC code possessing two classes of different error-correcting capabilities and allocate the flipping error on the strong part of the code bit.
A detailed description of the proposed system is shown in the following example.

\begin{figure}[hbt]
\begin{center}
\includegraphics[width=0.5\textwidth]{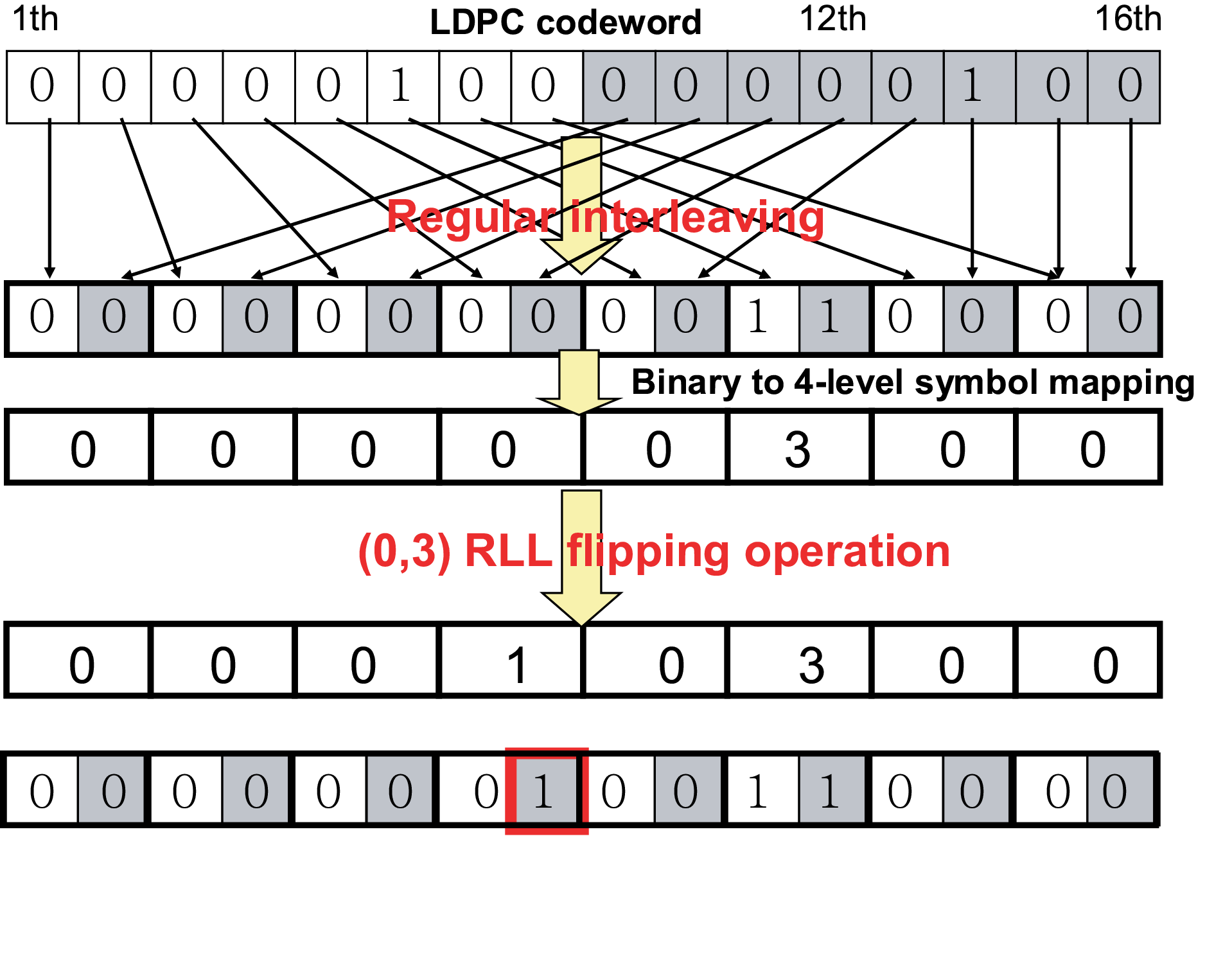}
\caption{Allocation technique type I for 4-level recording system under $(0,3)$ RLL constraint}
\label{uep2}
\end{center}
\end{figure}

 {\bf Example 1:} As illustrated in Fig.\,\ref{uep2} for the proposed allocation technique type I, a binary LDPC code sequence is $[0,0,0,0,0,1,0,0,0,0,0,0,0,1,0,0]$ and is interleaved to the sequence using (4.9) into $[0,0,0,0,0,0,0,0,0,0,1,1,0,0,0,0]$. Then, the binary sequence is mapped to the symbol sequence as $[0,0,0,0,0,3,0,0]$ using the natural mapping technique discussed in Section 4.3, and the symbol sequence is shifted to match the 4-level RLL (0,3) constraints as $[0,0,0,1,0,3,0,0]$. The 4th consecutive zeros is flipped to a non-zero symbol 1. The binary sequence with 4-level $(0,3)$ RLL constraint is consequently $[0,0,0,0,0,0,0,1,0,0,1,1,0,0,0,0]$. Hence, by setting the non-zero symbol for the RLL constraint to symbol 1 ($Z_1$=0, $Z_2$=1), the even-positioned symbols suffer from flipping errors, and the odd-positioned symbols are free of flipping errors. Therefore, the strong correcting bits are included in the invulnerable part from the $9th$ to the $16th$ position of the original LDPC code bit and have the RLL flipping error, and the weak correcting bits are contained in the vulnerable part from $1st$ to $8th$ position and are free from the RLL flipping error. It is also worth mentioning that RLL flipping symbol 1 ($Z_1$=0, $Z_2$=1) is a short-distance flipping at a distance of 0.8.

\subsection{Proposed UEP Scheme Type II}
  A UEP scheme type I is proposed for designing the strong part of the code bit with short-distance flipping and is designed with the consideration of reducing the burden of recovering the flipped error for the read side. However, the configuration of type I means that the strong code bit on the labeling needs to be set with a large number of nearest neighbors, which results in performance degradation according to the reasons discussed in Section III C. Based on these considerations, another consideration is introduced for type II in order to design the strong part of the code bit with long-distance flipping, such that the strong code bit for scheme type II is located where the labeling has a fewer number of $NN$s. The system block diagram is also illustrated in Fig.\,\ref{ueprxtx}. The difference is the bit ordering for type II, which is presented as
\begin{eqnarray}
\pi_{II}(l)=\left\{
            \begin{aligned}
            & \{M\cdot l+1 \}mod~(N+1),~0\leq l\leq N-2  \\
            &  l-1,~l=N-1,
            \end{aligned}
         \right.
\end{eqnarray}
where $\pi_{II}$ denotes the ordering operation of a type II regular interleaver.
The type II interleaver allocates the strong correction bit to the labeling bit with a fewer number of $NN$s in order to enhance the system performance.
 The proposed UEP Scheme type II is presented in the following example.

\begin{figure}[hbt]
\begin{center}
\includegraphics[width=0.5\textwidth]{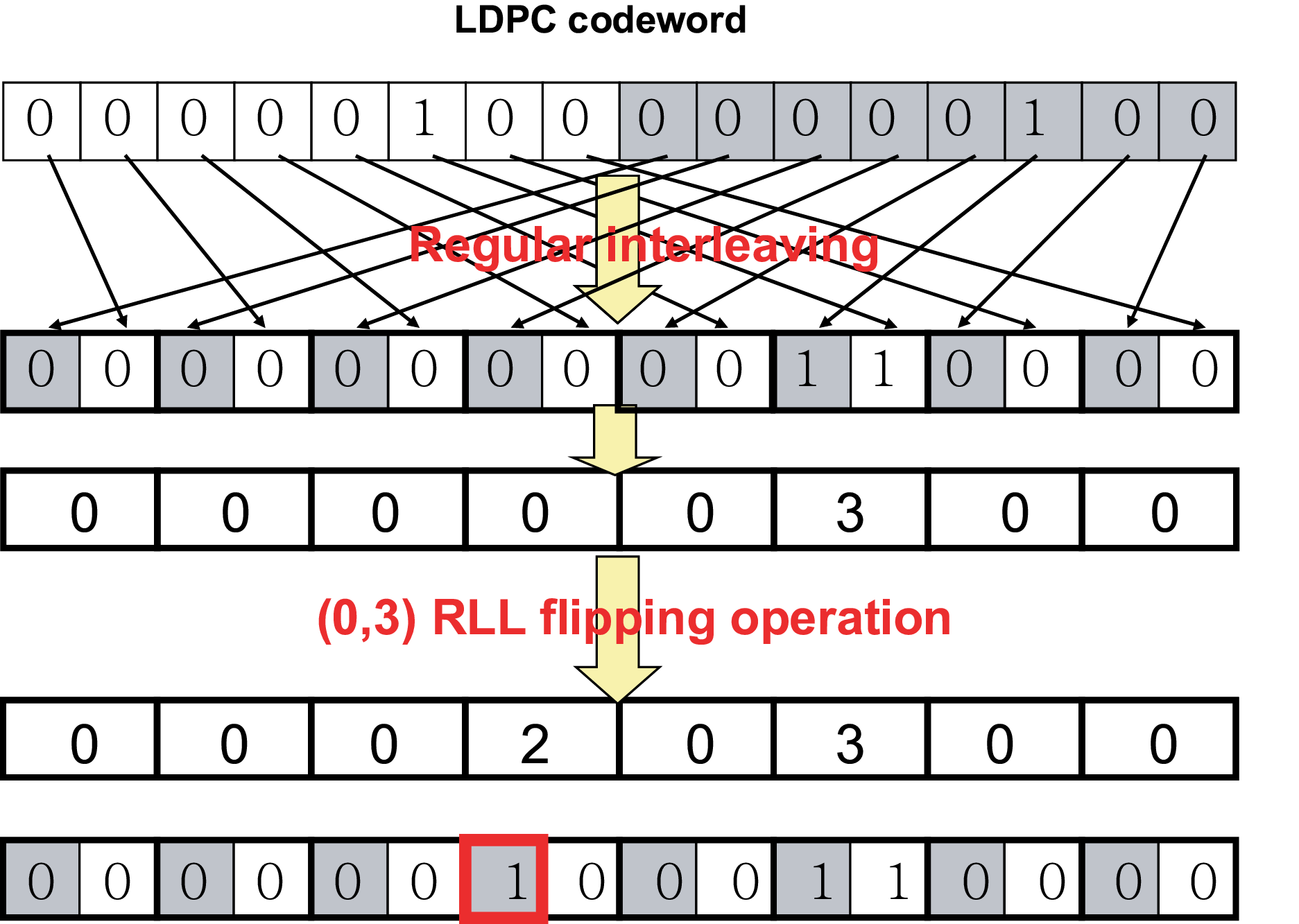}
\caption{Allocation technique for type II for a 4-level recording system under $(0,3)$ RLL constraints}
\label{uep3}
\end{center}
\end{figure}

  {\bf Example 4.2:} An example of the proposed allocation technique for type II is illustrated  in Fig.\,\ref{uep3}. The regular interleaver using (4.10) reverses the order compared to type I where the first half of the code bits are interleaved to even positions and the other half to the odd positions. Also, the non-zero symbol 2 ($Z_1$=1, $Z_2$=0) is set to match for RLL constraints. Therefore, the invulnerable part of the original LDPC code bit is similar to type I and suffers from RLL flipping errors, and the vulnerable part is free from RLL flipping errors. It is worth mentioning that the invulnerable part contains labeling with a fewer number of $NN$s. RLL flipping symbol 2 ($Z_1$=1, $Z_2$=0) is long-distance flipping with natural labeling. The scheme type II allocates the strong code bit to signal $Z_1$ with a fewer number of nearest neighbors for which $NN$($Z_1$)=1 and $NN$($Z_2$)=3.

 \subsection{Simulation Results}

PR channel $\overline{h}$=$(1,2,2,1)$ without jitter noise is evaluated as a theoretical experiment. In Fig.\,\ref{uep4} and the following figures, the outer iteration is set to $U_{o}=15$ and the inner iteration to $U_{i}=1$, where the non-reset LDPC decoder is used. Since the proposed UEP system requires two-level protection, we follow PEG construction \cite{peg} to construct possible LDPC codes. All trial LDPC code parameters of this section are summarized in Table \ref{ueptable2}. The node distribution is limited to two levels of error capability and leads to two or three different kinds of weight.

\begin{table*}[ht]\scriptsize
\centering
\caption{The node distribution for the UEP LDPC code used in this section.}
\begin{tabular}[b]{|c|c|c|c|c|c|c|}
\hline
UEP Code&Code length&Code rate&VND&$\delta$&CND&$\gamma$ \\
\hline
Code 1&4608&0.65&[2,3,5]&[0.442,0.0874,0.4706]&[10,11]&[0.96782,0.03218] \\
\hline
Code 2&4608&0.65&[2,5]&[0.5,0.5]&[10,11]&[0.9707,0.0293]  \\
\hline
Code 3&4608&0.65&[2,6]&[0.5,0.5]&[11,12]&[0.95937,0.04063]  \\
\hline
Code 4&4608&0.65&[2,7]&[0.5,0.5]&[12,13,14]&[0.8691,0.02416,0.10674] \\
\hline
Code 5&4608&0.65&[2,4]&[0.5,0.5]&[8,9,10]&[0.3289,0.66692,0.00418]\\
\hline
Code 6&4608&0.65&[3,4]&[0.5,0.5]&[10,11]&[0.96794,0.032]\\
\hline
Code 7&4608&0.65&[3,5]&[0.5,0.5]&[11,12]&[0.5169,0.4831]\\
\hline
Code 8&4608&0.75&[2,5]&[0.5,0.5]&[13]&[1]\\
\hline
Code 9&4608&0.8&[2,5]&[0.5,0.5]&[16,17,18]&[0.0163,0.9230,0.0607]\\
\hline
Code 10&11520&0.65&[2,5]&[0.5,0.5]&[9,10]&[0.00074,0.99926]\\
\hline
Code 11&11520&0.65&[2,6]&[0.5,0.5]&[10,11]&[0.85615,0.14385]\\
\hline
Code 12&11520&0.65&[2,7]&[0.5,0.5]&[10,11]&[0.713,0.287]\\
\hline
\end{tabular}
\label{ueptable2}
\end{table*}
For Fig.\,\ref{uep4}, irregular $(4608,3000)$ PEG-LDPC code rate 0.65 is used where the outer iteration $U_{o}=15$ and the inner iteration $U_{i}=1$. The proposed scheme type I and type II are compared. Here, Curve (A) to (D) show the results when proposed scheme Type I is applied, and the results when proposed scheme Type II is applied while are indicated by Curve (E) to (H). For the flipped systems represented by Curve (B), Curve (D), Curve (G) and Curve (H), PEG-LDPC code for $VND= [2,3,5]$, $[2,6]$, $[2,7]$ has a worse BER performance than $[2,5]$. We observe that the flipped system using UEP scheme type I allocates RLL flipping error to the signal $Z_2$ with a strong code bit resulting in a BER performance that is close to the non-flipped system. However, signal $Z_2$ has more nearest neighbors than signal $Z_1$. Hence, Curve (E) and Curve (F) reveal that the BER performance is enhanced by a strong code bit with less nearest neighbor labeling, and the flipped system using UEP scheme type II also performs better than UEP scheme type I.
From this result, it can be concluded that, although short-distance flipping lowers the degradation of the flipped bit and Curve (C) is very close to Curve (D), scheme type II has a better UEP capability and performs better than type I when applied to a non-flipped system. Hence, Curve (F) reveals the best performance for the flipped system.
\begin{table*}[ht]\scriptsize
\centering
\caption{The node distribution for the UEP LDPC codes used in Fig.\,\ref{uep4}.}
\begin{tabular}[b]{|c|c|c|c|c|c|c|}
\hline
UEP Code&Code Length&Code Rate&VND&$\delta$&CND&$\gamma$ \\
\hline
Code 1&4608&0.65&[2,3,5]&[0.442,0.0874,0.4706]&[10,11]&[0.96782,0.03218] \\
\hline
Code 2&4608&0.65&[2,5]&[0.5,0.5]&[10,11]&[0.9707,0.0293]  \\
\hline
Code 3&4608&0.65&[2,6]&[0.5,0.5]&[11,12]&[0.95937,0.04063]  \\
\hline
Code 4&4608&0.65&[2,7]&[0.5,0.5]&[12,13,14]&[0.8691,0.02416,0.10674] \\
\hline
\end{tabular}
\label{ueptable3}
\end{table*}
\begin{figure}[hbt]
\includegraphics[width=0.5\textwidth]{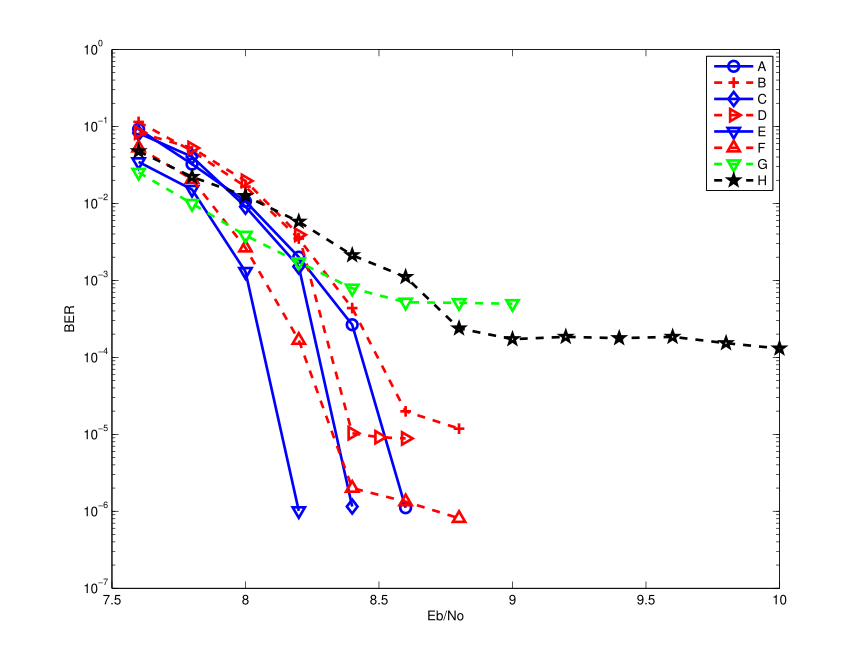}
\captionsetup{font=scriptsize}
\caption{BER results for a 4-level $(0,3)$ RLL constraint over PR channel $\overline{h}$=$(1,2,2,1)$.
 (A) Non-flipped system type I using UEP LDPC Code 1;
 (B) Flipped system type I using UEP LDPC Code 1;
 (C) Non-flipped type I system using UEP LDPC Code 2;
 (D) Flipped system type I using UEP LDPC Code 2;
 (E) Non-flipped system type II using UEP LDPC Code 2;
 (F) Flipped system type II using UEP LDPC Code 2;
 (G) Flipped system type II using UEP LDPC Code 3;
 (H) Flipped system type II using UEP LDPC Code 4.
  }
\label{uep4}
\end{figure}

\begin{figure}[hbt]
\includegraphics[width=0.5\textwidth]{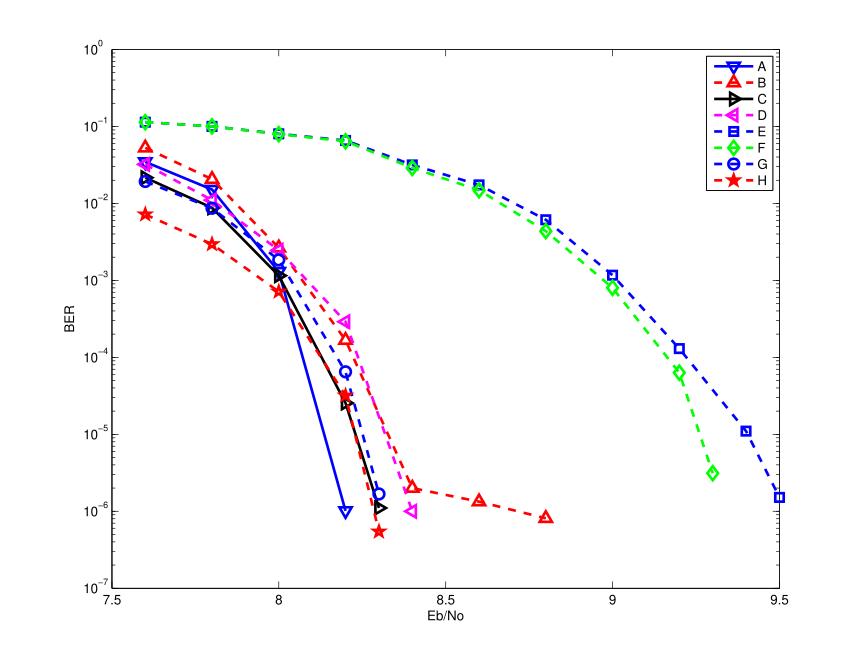}
\captionsetup{font=scriptsize}
\caption{BER results for a 4-level $(0,3)$ RLL constraint over PR channel $\overline{h}=(1,2,2,1)$ using proposed scheme type II.
 (A) Non-flipped system using $U_{o}=15$ and $U_{i}=1$;
 (B) Flipped system using $U_{o}=15$ and $U_{i}=1$;
 (C) Non-flipped system using $U_{o}=5$ and $U_{i}=3$;
 (D) Flipped system using $U_{o}=5$ and $U_{i}=3$;
 (E) Flipped system using $U_{o}=1$ and $U_{i}=15$;
 (F) Flipped system using $U_{o}=1$ and $U_{i}=75$;
 (G) Flipped system using $U_{o}=15$ and $U_{i}=5$;
 (H) Flipped system using $U_{o}=15$ and $U_{i}=15$.
  }
\label{uep5}
\end{figure}

Now, the focus is on increasing the inner LDPC iteration to enhance the UEP capability. By adjusting the parameters for $U_{o}$ and $U_{i}$ illustrated in Fig.\,\ref{uep5}, Curve (E) and Curve (F) show that $U_{o}=1$, and the only processing a single inner iteration of the LDPC decoder is not effective for a flipped system and does not solve RLL flipping error. For Curve (A) and Curve (B), the error floor for the flipped system is displayed after 8.4dB. Hence, by setting $U_{o}=5$ and $U_{i}=3$, an identical overall iteration of $U_{o}=15$ and $U_{i}=1$ is produced. Although Curve (C) for the non-flipped system using $U_{o}=5$ and $U_{i}=3$ shows a slightly inferior result to that of the Curve (A) for the non-flipped system using $U_{o}=15$ and $U_{i}=1$, by less than 0.1dB, however Curve (C) and Curve (D) show that flipping error is successfully eliminated by the strong code bit and no error floor is present. The flipped system has an excellent performance that is close to the non-flipped system by only 0.1dB. The flipped system illustrated by Curve (D) where $U_{o}=5$ and $U_{i}=3$ is nearly convergent in the number of iterations compared to  Curve (H) where $U_{o}=15$ and $U_{i}=15$, since Curve (H) is separated from Curve (D) by 0.1dB.  Hence, for a 4608-length code with rate 0.65, code parameter $VND=[2,5]$ with $\lambda_{k}$=[0.5,0.5] and $CND=[10,11]$ with $\gamma=[0.9707,0.0293]$ is a recommended code.

\begin{figure}[hbt]
\includegraphics[width=0.5\textwidth]{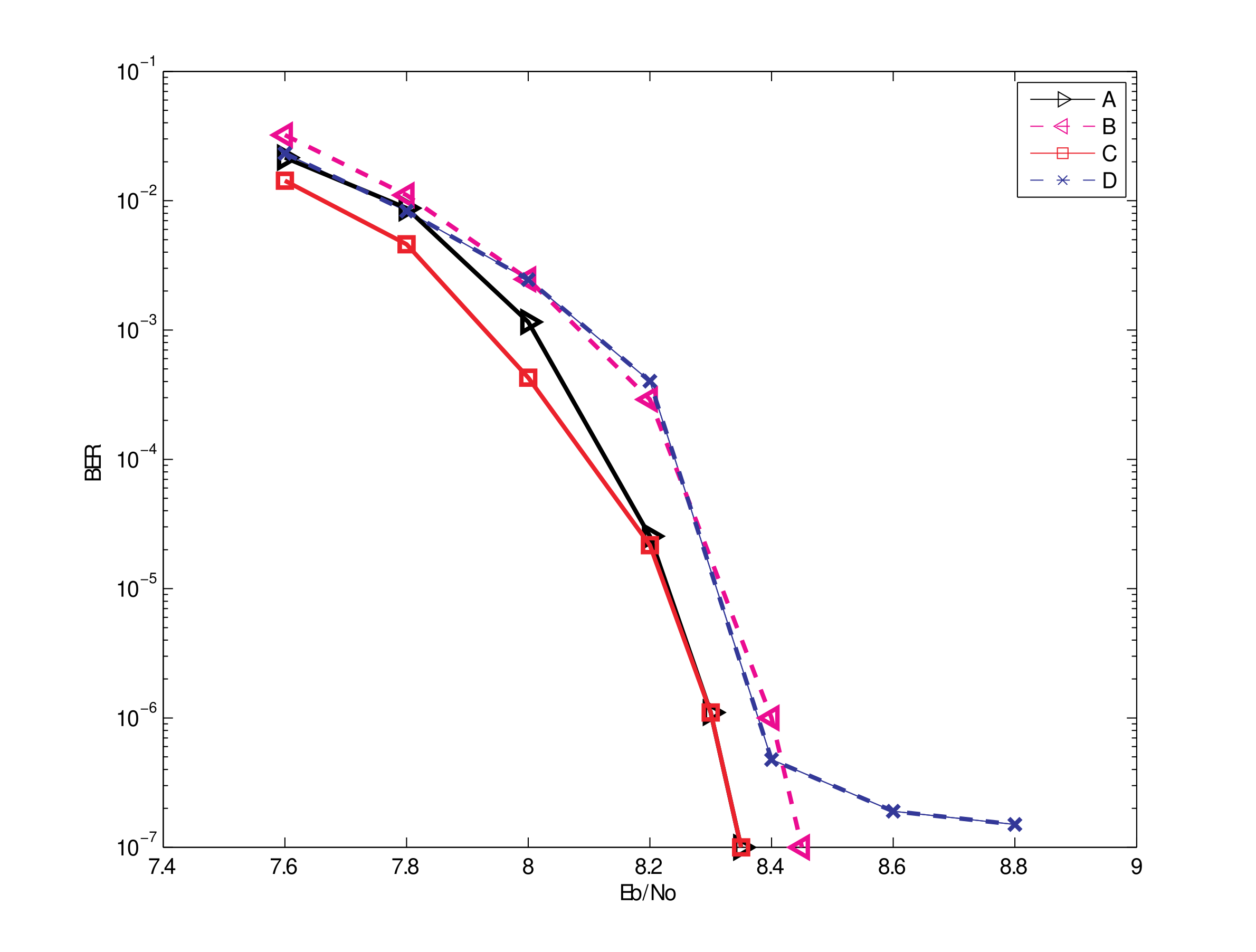}
\captionsetup{font=scriptsize}
\caption{BER results for a 4-level $(0,3)$ RLL constraint over PR channel $\overline{h}=(1,2,2,1)$ using PEG-LDPC code where $U_{o}=5$ and $U_{i}=3$ (rate-0.65 code).
 (A) Non-flipped system using natural mapping;
 (B) Flipped system using natural mapping;
 (C) Non-flipped system using Gray mapping;
 (D) Flipped system using Gray mapping.
  }
\label{uep13}
\end{figure}

It is interesting to compare the system performance of the signal labeling presented in Figure 5. As illustrated in Fig.\,\ref{uep13}, it can be observed that the Gray mapping improves the nearest neighbors for the weak ECC part compared to natural mapping, which leads to the comparison of Curve (A) and Curve (C) where the performance is slightly better over the low SNR region, and the performance over the high SNR region is identical. However, the long-distance flipping for Gray mapping is equal to 7.2, which results in a burden for the decoder when attempting to recover the flipping error and reveals an error floor of $10^{-7}$ in Curve (D). This simulation result is identical to the predicted performance of the system design discussed in Section IV C. Consequently, the proposed scheme Type II provides a better flipping error solution, which compensates for the inferior property of the nearest neighbor of the weak code bit. Finally, we present the merit of the proposed system using a suitable signal labeling design.

\begin{figure}[hbt]
\includegraphics[width=0.5\textwidth]{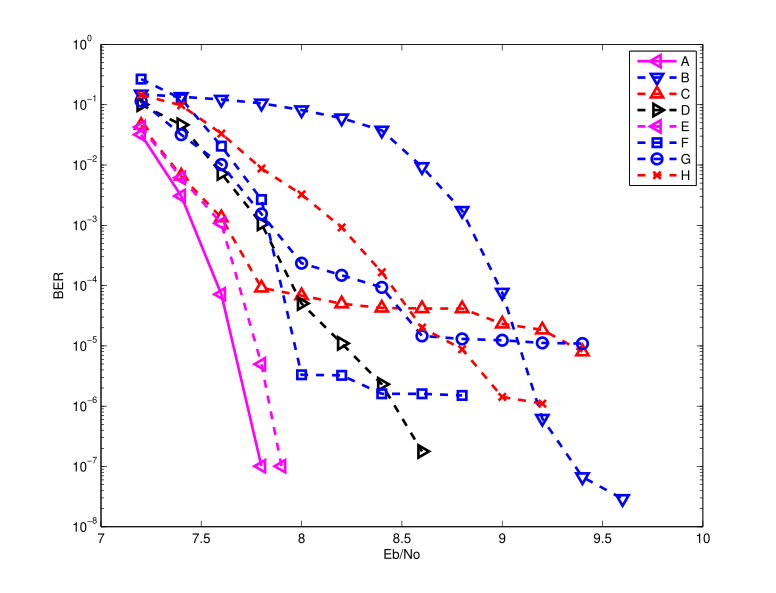}
\captionsetup{font=scriptsize}
\caption{BER results for a 4-level $(0,3)$ RLL constraint over PR channel $\overline{h}=(1,2,2,1)$  using proposed scheme type II.
 (A) Non-flipped system using $U_{o}=5$ and $U_{i}=5$;
 (B) Flipped system using $U_{o}=1$ and $U_{i}=15$;
 (C) Flipped system using $U_{o}=5$ and $U_{i}=3$;
 (D) Flipped system using $U_{o}=3$ and $U_{i}=5$;
 (E) Flipped system using $U_{o}=5$ and $U_{i}=5$;
 (F) Flipped system using $U_{o}=15$ and $U_{i}=1$;
 (G) Flipped system using $U_{o}=5$ and $U_{i}=5$;
 (H) Flipped system using $U_{o}=5$ and $U_{i}=5$.
  }
\label{uep6}
\end{figure}

A Long LDPC code length has a great number of RLL flipping errors from a statistical sense, but the proposed flipped system using UEP scheme type II can still solve RLL flipping errors while simultaneously increasing the ECC ability and the number of RLL flipping errors. Fig.\,\ref{uep6} illustrates a long code length of (11520,7488) PEG-LDPC code with rate $0.65$. For Curve (A) to (F), Code 10 is used, for Curve (G), Code 11 is used, and for Curve (H), Code 12 is used. The BER performance for $VND=[2,5]$ using $U_{o}=5$ and $U_{i}=5$ is still better than $[2,6]$ and $[2,7]$, as illustrated by Curve (E), Curve (G) and Curve (H). Iterative decoding where $U_{o}=1$ is not effective for the flipped system, where Curve (B) and Curve (J) using $U_{o}=1$ and $U_{i}=15$ displays an error floor region. Curves (B) to (F) show different $U_{o}$ and $U_{i}$ values under an identical number of overall iterations, $U_{o}\times U_{i}=15$. In particular, only Curve (D) using $U_{o}=3$ and $U_{i}=5$ solves the flipping error. Moreover, Curve (E) using $U_{o}=5$ and $U_{i}=5$ reveals a better BER performance for the flipped system. For Curve (A), a non-flipped system using $U_{o}=5$ and $U_{i}=5$, is close to the flipped system of Curve (E) separated by only 0.1dB. Hence, code parameter $VND=[2,5]$ with $\lambda_{k}=[0.5,0.5]$ and $CND=[9,10]$ with $\gamma=[0.00074,0.99926]$ is a recommended code. The result indicates that check node with a wight of 10 produces a suitable check code. This property will be investigated in the next section.
\begin{table*}[ht]\scriptsize
\centering
\caption{The node distribution for the UEP LDPC codes used in Fig.\,\ref{uep6}.}
\begin{tabular}[b]{|c|c|c|c|c|c|c|}
\hline
UEP Code&Code Length&Code Rate&VND&$\delta$&CND&$\gamma$ \\
\hline
Code 10&11520&0.65&[2,5]&[0.5,0.5]&[9,10]&[0.00074,0.99926]\\
\hline
Code 11&11520&0.65&[2,6]&[0.5,0.5]&[10,11]&[0.85615,0.14385]\\
\hline
Code 12&11520&0.65&[2,7]&[0.5,0.5]&[10,11]&[0.713,0.287]\\
\hline
\end{tabular}
\label{ueptable4}
\end{table*}

\begin{figure}[hbt]
\includegraphics[width=0.5\textwidth]{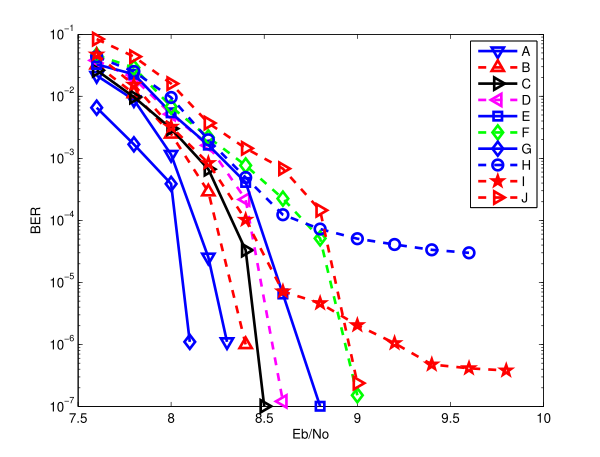}
\captionsetup{font=scriptsize}
\caption{BER results for a 4-level $(0,3)$ RLL constraint over PR channel $\overline{h}=(1,2,2,1)$ using PEG-LDPC code.
 (A) Non-flipped system using $U_{o}=5$ and $U_{i}=3$ (rate-0.65 code);
 (B) Flipped system using $U_{o}=5$ and $U_{i}=3$ (rate-0.65 code);
 (C) Non-flipped system using $U_{o}=5$ and $U_{i}=3$ (rate-0.75 code);
 (D) Flipped system using $U_{o}=5$ and $U_{i}=3$ (rate-0.75 code);
 (E) Non-flipped system using $U_{o}=5$ and $U_{i}=5$ (rate-0.8 code);
 (F) Flipped system using $U_{o}=5$ and $U_{i}=5$ (rate-0.8 code);
 (G) Non-flipped system using $U_{o}=5$ and $U_{i}=3$ (rate-0.65 code);
 (H) Flipped system using $U_{o}=5$ and $U_{i}=3$ (rate-0.65 code);
 (I) Flipped system using $U_{o}=5$ and $U_{i}=3$ (rate-0.65 code);
 (J) Flipped system using $U_{o}=5$ and $U_{i}=3$ (rate-0.65 code).
  }
\label{uep8}
\end{figure}

As the code rate of the system is increased, the corresponding capability becomes weaker, the results for which are illustrated in Fig.\,\ref{uep8}, where the proposed scheme type II is applied. For Curve (A) and (B), (4608,3000) LDPC Code 2 with a rate 0.65 is used. For Curve (C) and (D), (4608,3456) LDPC Code 8 with a rate 0.75  is used. For Curve (E) and (F), (4608,3686) LDPC Code 9 with a rate 0.8 is used. For Curve (G) and (H), regular-(4608,3000) and code parameter $VND=[3]$ $CND=[9]$. LDPC code with a rate 0.65 is used. For Curves (I) and (J), irregular-(4608,3000) LDPC Code 6 with a rate 0.65 and Code 7 with a rate 0.75 are used respectively.  Curve (C) and Curve (D) show that the flipped system using  $(4608,3456)$ LDPC code rate 0.75 is close to the non-flipped system by 0.1dB. For the higher code rate of 0.8, Curve (E) and Curve (F) also illustrate satisfactory performance. In order to compare the performance with regular LDPC code, Curve (G) shows the powerful error-correcting capability over irregular code of the non-flipped system. However, an error floor region is revealed in the flipped system illustrated by Curve (H), even though the non-flipped system performance is better than irregular codes. Consequently, a number of the good LDPC code ensembles are irrelevant for flipped systems. We also present the code performance for $VND=[3,4]$ in Curve (I), which results in an error floor, since the UEP capability is not sufficient. Moreover, Curve (J) using code parameter $VND=[3,5]$ reveals no error floor and is therefore presented as the recommended code.
\begin{table*}[ht]\scriptsize
\centering
\caption{The node distribution for the UEP LDPC codes used in Fig.\,\ref{uep8}.}
\begin{tabular}[b]{|c|c|c|c|c|c|c|}
\hline
UEP Code&Length&Rate&VND&$\delta$&CND&$\gamma$ \\
\hline
Code 2&4608&0.65&[2,5]&[0.5,0.5]&[10,11]&[0.9707,0.0293]  \\
\hline
Code 6&4608&0.65&[3,4]&[0.5,0.5]&[10,11]&[0.96794,0.032]\\
\hline
Code 7&4608&0.65&[3,5]&[0.5,0.5]&[11,12]&[0.5169,0.4831]\\
\hline
Code 8&4608&0.75&[2,5]&[0.5,0.5]&[13]&[1]\\
\hline
Code 9&4608&0.8&[2,5]&[0.5,0.5]&[16,17,18]&[0.0163,0.9230,0.0607]\\
\hline
\end{tabular}
\label{ueptable5}
\end{table*}

\begin{figure}[hbt]
\captionsetup{font=scriptsize}
\includegraphics[width=0.5\textwidth]{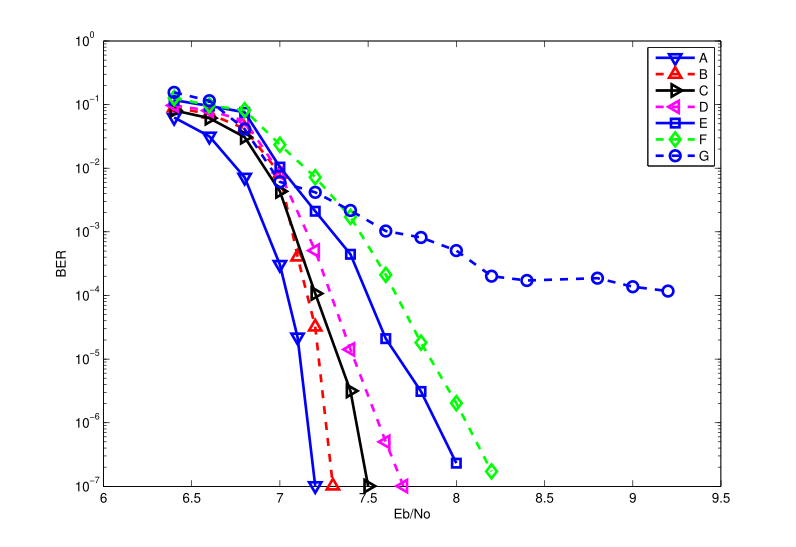}
\caption{BER results for 4-level $(0,3)$ RLL constraint over optical recording channel using PEG-LDPC code, where $\beta$ = 0.15. The proposed scheme type II is applied.
 (A) Non-flipped system using $U_{o}=5$ and $U_{i}=3$ (rate-0.65 code);
 (B) Flipped system using $U_{o}=5$ and $U_{i}=3$ (rate-0.65 code);
 (C) Non-flipped system using $U_{o}=5$ and $U_{i}=3$ (rate-0.75 code);
 (D) Flipped system using $U_{o}=5$ and $U_{i}=3$ (rate-0.75 code);
 (E) Non-flipped system using $U_{o}=5$ and $U_{i}=5$ (rate-0.8 code);
 (F) Flipped system using $U_{o}=5$ and $U_{i}=5$ (rate-0.8 code);
 (G) Flipped system using $U_{o}=5$ and $U_{i}=3$ (rate-0.65 code).
  }
\label{uep9}
\end{figure}

In order to consider a more practical channel, jitter noise is applied in Fig.\,\ref{uep9}. The result is similar to the theoretical channel. For Curve (A) and (B), (4608,3000) LDPC Code 2 with a rate 0.65 is used. For Curve (C) and (D), (4608,3456) LDPC Code 8 with a rate 0.75  is used. For Curve (E) and (F), (4608,3686) LDPC Code 9 with a rate 0.8 is used. For Curve (G), (H) and (I) regular-(4608,3000) LDPC code with rate 0.65 and code parameter $VND=[3]$ $CND=[9]$, irregular-(4608,3000) LDPC Code 6 with rate 0.65 and Code 7 are used respectively. The result is similar to the theoretical recording channel over LDPC code rate 0.65 and the flipped system is close to the non-flipped system 0.1dB. Curve (C) and Curve (D) show that the flipped system using  $(4608,3456)$ LDPC code rate 0.75 is close to the non-flipped system 0.2dB. For a higher code rate of 0.8, Curve (E) and Curve (F) also show the the error floor region is eliminated. Curve (G) also shows the error floor region in the flipped system.

$~~~~~~~$To begin with traditional multilevel coded modulation for unequal protection, the system performance of signal labeling using Ungerboeck partitioning and block partitioning is discussed. In order to maintain the UEP property of different levels, reducing the number of nearest neighbors ($NN$) at every partition level is necessary to be considered. Since multilevel error protection is required for the UEP-coded system, the conventional signal set partitioning for maximizing intraset distance at each level is not enough for constructing modulation codes. Hence, multilevel codes using block partitioning provide a large decrease in effective error coefficients in the first several decoding stages and perform well in the UEP system.

A multilevel recording system provides a space to improve the deliberate flipping system. We implement a 4-level recording system applying UEP technique to recover the flipped bit at the read side. We illustrate UEP property on both LDPC code and signal labeling configuration to construct the proposed technique. Two types of flipped bits with different Euclidean distances provide different flipped-bit interference over the recording system. However, the error coefficient term affects system performance which is caused by a large hamming distance code bit corresponding to signal labeling along with less number of nearest neighbors. Hence, the proposed UEP scheme type II is designed under this consideration and reveals a satisfactory performance. Iterative decoding possesses inner iteration and outer iteration between the MAP equalizer and LDPC decoder. Under identical decoding complexity, increasing inner iteration allows to the enhancement of the UEP capability. Hence, the error floor from the flipped bit is alleviated. We compare the BER performance of several possible UEP LDPC codes for the proposed system and recommend an appropriate LDPC node distribution for the flipped-bit system. It is worth mentioning that although the non-flipped bit system using regular-LDPC code with medium block length has better performance than irregular-LDPC code, the error correcting capability is still insufficient for recovering flipped-bit.

\section{Optimization Approach for Proposed Flipped System using Density Evolution and Differential Evolution}

Density evolution \cite{Richardson} (DE) is a general method for determining the capacity of LDPC codes by referring to the evolution of the probability density functions (pdfs) of the messages being passed around in the iterative decoder. Such Knowledge of the pdfs allows one to predict under which channel conditions (e.g., SNRs) the decoder bit error probability will converge to zero. As the codeword length tends to infinity, the same ensemble of a random code will be more and more likely to be cycle-free. Under the cycle-free assumption, we can analyze the decoding algorithm straight-forwardly because incoming messages to every node are independent. In the two-stage computation, for the variable and check node stage, without loss of generality, the assumption is made that the all-0 codeword was sent. The DE algorithm possesses the symmetry condition to ensure the probability of error of a message is $0$ and the probability of error at the $l$-th iteration is a non-increasing function of $l$. This property leads to a stability condition that controls the behavior of density evolution near the zero probability of error.

  In \cite{Richard}, the optimized degree distribution from DE is provided and DE property is discussed. The stability condition which ensures the convergence of the density evolution for mutual information should be close to one. The stability condition gives an upper bound on the number of degree-2 variable nodes. Then, the concentration theorem is proven to ensure all codes in an ensemble have the same behavior. Hence, the performance prediction of a specific long code is possible via the DE algorithm. Particularly, long code possesses a decoding threshold. The threshold separates one region for reliable communication and another region does not. DE allows determining the decoding threshold of an LDPC code ensemble and its gap from Shannon's limit. With channel symmetry as one of its fundamental assumptions, DE has been widely and successfully applied to different channels, including binary erasure channels (BECs), binary symmetric channels (BSCs), and binary additive white Gaussian noise (BiAWGN) channels. In the sense of obtaining a minimum threshold, DE also allows to do the optimization of irregular LDPC degree distribution. For the ISI channel, Kavcic \cite{Kavcic} provides an investigation for the performance limit of the decoder and demonstrates the potential for trading off the computational requirements. The maximum achievable information rate for the scheme of MAP equalizer and LDPC decoder is proven. Following the thread in \cite{Kavcic}, the concentration theorem must also hold for the proposed flipped system. We organize as following two theorems.

  {\bf Theorem 1 :} Let $s$ be the transmitted codeword with code length $n$. The $Z^{(u)}(s)$ is the random variable that represents the number of RLL flipped messages after $u$ rounds of variable-to-check message-passing decoding algorithm. The code graph is chosen uniformly at random from the ensemble of graphs with degree polynomials $\lambda(x)$ and $\rho(x)$. Let $n_{e}$ denote the number of variable-to-check edges in the graph. For an arbitrarily small constant $\varepsilon>0$, there exists a positive number $\beta$, such that if $n>\frac{2\gamma}{\varepsilon}$, then
   \begin{equation}
 Prob(|\frac{Z^{(u)}(s)}{n_{e}}-p_{s}^{u}|\geq\varepsilon)\leq e^{-\beta \varepsilon^{2}n}
 \end{equation}
 where the $p_{s}^{u}$ is the error concentration probability denoted as the error probability of tree-like
 decoding under all possible tree types while the codeword $s$ is transmitted. The detailed description is shown in \cite{Kavcic}.

  {\bf Proof :} The proof is identical in \cite{Kavcic} and mainly apply the Azuma's inequality.

  {\bf Theorem 2 :} There exists a linear block code $C$ in the ensemble $C( \lambda(x), \rho(x))$ of LDPC code set, such that for any variance $\sigma< \sigma^{*}$, $\sigma^{*}$ denoted as the decoding threshold, the probability of error can be made arbitrarily low. The $Z^{(u)}$ denotes the number of RLL flipped messages after $u$ rounds of variable-to-check message-passing decoding algorithm on the joint channel/code graph of the code $C$, the following inequality must hold.
   \begin{equation}
 Prob(\frac{Z^{(u)}}{n_{e}}\geq 2\varepsilon| C( \lambda(x), \rho(x)))\leq 4e^{-\beta' \varepsilon^{2}n}
 \end{equation}

  {\bf Proof :} Since the random variable of the flipped bit satisfies the property of independent uniform distribution, the proof is also identical in \cite{Kavcic}.

Since the flipped bit is limited to a segment of the code bit, there are a number of hard errors located on the flipped part of the code bit. Hence, the error probability of the flipped part denoted as $p_{f}^{u}$ is necessarily small at a certain level. In order to specify the condition of error floor cancelation, the ratio of the error probability of the non-flipped part and flipped part denoted as $\psi=p_{nf}^{u}/p_{f}^{u}$ is greater than a threshold value $\eta_e$. Based on Theorem 1, there are two viewpoints for the proposed flipped system. (i) Equation (5.1) presents the decoding behavior of the flipped bit. The error floor allows canceling in exponential decay while for an arbitrarily small constant $\varepsilon>0$; (ii) The threshold value $\eta_e$ is related to the positive number $\beta$, where depends only on $\delta$, $\gamma$, and $u$. It seems a relation for $\beta$ such that $\beta$ is a function of $\eta_e$ denoted as $\beta=f(\eta_e)$. In order to maintain a high threshold value $\eta_e$, the weight of $\delta$ in the flipped part is designed to be relatively larger than the non-flipped part.

  For practical application of code search, we do not apply the DE to optimize the LDPC node distribution directly, for the following reasons. But we still allow it to do the code performance evaluation.
  Thus, the infinite-length cycle-free assumptions rely on DE to determine the optimal degree distribution. However, when optimal degree distributions are adopted in the design of short- or medium-length codes, these codes are generally susceptible to high error floors due to short cycles and trapping sets. In particular, the presence of cycles involving mostly, or only, degree-2 variable nodes deteriorates the LDPC decoder. This phenomenon becomes worse while applying to the flipped system.

\begin{figure}[hbt]
\centering
\captionsetup{font=scriptsize}
\includegraphics[width=0.5\textwidth]{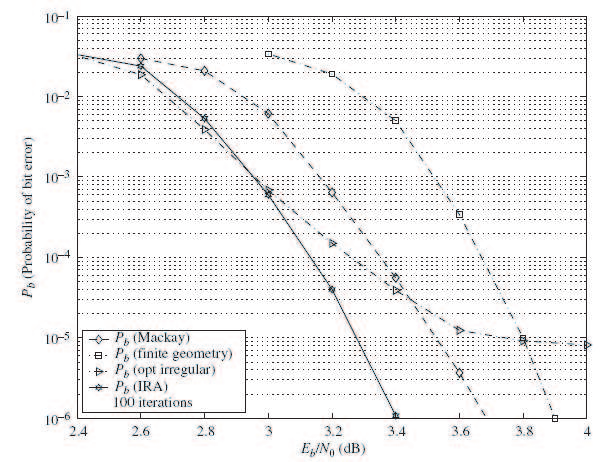}
\caption{Comparison of four length-4161, rate-0.82 LDPC codes, including a near-optimal ("opt" in the figure) irregular one that displays a high error-rate floor in \cite{YangRyan}.
  }
\label{de1}
\end{figure}

  Consider the design of a (4161,3430)-LDPC code with a rate of 0.82 and compete
  with other known codes with those parameters. Near-optimal degree distributions
  with $d_{v} = 8$ and $d_{c} = 20$ were found \cite{YangRyan} as follows.
  \begin{eqnarray}
   \lambda(X)& = &0.2343X + 0.3406X^{2} + 0.2967X^{6} + 0.1284X^{7},\nonumber\\
   \rho(X)& = &0.3X^{18} + 0.7X^{19}.
  \end{eqnarray}
  The number of degree-2 variable nodes would be computed as $1685$.
  The maximum number of degree-2 variable nodes possible before a cycle involving only these degree-2 nodes is created to limit the girth of this code to no greater than 10. Such graphical configurations give rise to an error floor. In \cite{YangRyan}, Fig.\,\ref{de1} presents that the irregular code with the near-optimal degree distribution in
(5.3) does indeed appear to have the best decoding threshold, but it is achieved at the expense of the error floor due to its large number of degree-2 variable nodes and their associated cycles. Hence, the code performance is necessary to be evaluated by Monte Carlo simulation for practical application.

\subsection{A Flipped and Non-flipped Based Density Evolution for Proposed UEP Recording System}
 Since the occurrence of a flipped bit is difficult to determine in a certain relationship or model, the proposed density evolution is designed to evaluate the code distribution for the flipped system.
 In this section, we derive the density evolution formulas that the proposed system is derived based on the Monte Carlo approach.
 Based on the simulation outlined in Section III E, it has been shown that the UEP LDPC code
 with $VND=[2,5]$ provides a better BER performance using BP decoding.
 DE allows evaluation of the node distribution for the proposed UEP scheme.
 In order to analyze the read side, iterative decoding between the MAP equalizer and the LDPC decoder
 is considered so as to perform density evolution. However, since the trellis evolution
 with closed-form solution for the MAP equalizer is unknown \cite{Kavcic}, we calculate numerically using
 Monte Carlo techniques for only one iteration of the MAP equalizer.
 For the sum-product-based DE, the complicated check node computation is presented as follows:
 \begin{equation}
 p_{u}^{(c)}=\Gamma^{-1}[(\Gamma[p_{u}^{(v)}])^{\ast(d_{c}-1)}].
 \end{equation}
 The operation $\Gamma$ corresponds to the change of density due to the hyperbolic tangent transformation $\phi$ function as occurs in the computation for the check node. We observe that the numerical computation $\Gamma$
 is difficult to implement. Although check node computation can easily be obtained through a table lookup \cite{ChungUrbanke}, it is still not feasible for the proposed system. Thus, we do not apply this approach and the computation is redirected to a simpler min-sum decoding algorithm.
 The min-sum based DE algorithm was applied in \cite{Jinghu}
 to analyze the BP-based min-sum decoding algorithm. The sum-product-based DE is substituted into the min-sum DE algorithm for two reasons: (i) The focus of this thesis is based on the LDPC code distribution for the flipped system rather than searching for the decoding threshold approaching channel capacity. Hence, the performance degradation of min-sum decoding can be neglected for the purpose of evaluating code distribution; (ii) In order to simplify the analysis process and without losing the functionality of identifying good codes, the min-sum algorithm is a good candidate for replacing the complicated sum-product based DE algorithm for evaluating an
 ensemble LDPC code performance. The min-sum-based
 density evolution for the proposed scheme is derived in the following section.
\subsection{Proposed Min-Sum Density Evolution Algorithm for UEP Recording System}
 For the RLL flipped system, the flipping error on the write side is dependent
 on a random LDPC codeword. Hence, transmitting all-zero or all-one codeword
 in the conventional DE algorithm can not be applied to a
 symbol-dependent system. Consequently, Monte Carlo technique is applied to the system using deliberate RLL flipping.
 The error message is represented by the bit error while simulating the decoding procedure. We define the error message corresponding to the variable node with $v_{i}\neq v'_{i}$, for $i=1,...,N$, where $\overline{v}$=(v$_{1}$,v$_{2}$,...,v$_{N}$) is the transmitted sequence and $\overline{v}'$=(v'$_{1}$,v'$_{2}$,...,v'$_{N}$) is the received sequence.
 Then, we simulate an error message while the first
 iteration passes from variable node $i$, for $i=1,2,...,N$,
 to check node $j$, for $j=1,2,...,M$, denoted as $\textbf{V}=[V_{1}, V_{2},..., V_{d_{c}}]$.
 Based on the simulation, the probability density function (pdf) of the error message denoted as $P^{(0)}_{\textbf{V}}(v)$ can be measured from $L_{d}(v_{i})$ at the input
 of the LDPC decoder illustrated in Fig.\,\ref{ueprxtx}. According to the proposed
 UEP scheme, $\textbf{V}$ can be divided into $\textbf{V}^{f}$ with  flipping errors
 and $\textbf{V}^{nf}$ without flipping errors. Consequently, the differences from the conventional DE are that one is the pdf calculation of the error message and the other is the transmission of the random codeword.

 Then, we derive the min-sum DE for the proposed scheme is derived. While processing during the $(u-1)$-th iteration,
 the probability of $\textbf{V}^{f}$ with magnitude greater
 than $x$ is denoted as

 \begin{equation}
 \phi_{+}^{f}(x)=\int_{x}^{\infty}P^{(u-1)}_{V^{f}}(v^{f})dv^{f}.
 \end{equation}
 and
 \begin{equation}
 \phi_{-}^{f}(x)=\int^{-x}_{-\infty}P^{(u-1)}_{V^{f}}(v^{f})dv^{f}.
  \end{equation}
 For $\textbf{V}^{nf}$,
 \begin{equation}
 \phi_{+}^{nf}(x)=\int_{x}^{\infty}P^{(u-1)}_{V^{nf}}(v^{nf})dv^{nf}.
  \end{equation}
 and
 \begin{equation}
 \phi_{-}^{nf}(x)=\int^{-x}_{-\infty}P^{(u-1)}_{V^{nf}}(v^{nf})dv^{nf}.
 \end{equation}
 are obtained.
  Consequently, the probability of error message $L$ from
 check node $j$ to variable node $i$ can be determined as for $l<0$,
  \begin{equation}
  \begin{split}
 Pr(L<l)= Pr\{odd~ number~ of~ negative~\\ value~ in~ V~ and~ |V_j|>|l|\}.
 \end{split}
  \end{equation}
  and for $l\geq0$,
  \begin{equation}
    \begin{split}
 Pr(L<l)=Pr\{even~ number~ of~ negative~\\ value~ in~ V,~ and~ |V_j|>|l|\}.
  \end{split}
  \end{equation}
 The probability distribution function of $L$ for $l<0$ is based on the Bernoulli equation,
\begin{equation}
\begin{split}
F_{L}(l)
=\sum^{d_{c}-1}_{k=1 k\in odd} {d_{c}-1\choose k}\sum_{v=1w=1}^{k}(\phi_{-}^{nf}(l))^{v}(\phi_{-}^{f}(l))^{k-v}\\\phi_{+}^{nf}(l))^{w}(\phi_{+}^{f}(l))^{k-w} \\
=\frac{1}{2}(\phi_{-}^{nf}(l)+\phi_{-}^{f}(l)+\phi_{+}^{nf}(l)+\phi_{+}^{f}(l))^{d_{c}-1}. ~~~~~~~~   
\end{split}
\end{equation}
Also for $l\geq0$,
\begin{equation}
\begin{split}
F_{L}(l)
=1- \sum^{d_{c}-1}_{k=1 k\in even} {d_{c}-1\choose k}\sum_{v=1w=1}^{k}(\phi_{-}^{nf}(l))^{v}\\(\phi_{-}^{f}(l))^{k-v}(\phi_{+}^{nf}(l))^{w}(\phi_{+}^{f}(l))^{k-w}. \nonumber\\
=\frac{1}{2}(\phi_{-}^{nf}(l)+\phi_{-}^{f}(l)-\phi_{+}^{nf}(l)-\phi_{+}^{f}(l))^{d_{c}-1}. ~~~~~~~~~
\end{split}
\end{equation}
By taking the derivative of $F_{L}(l)$ with respect to $l$, we have for $l\geq0$
\begin{equation}
\begin{split}
Q_{L}^{(u)}(l) = \frac{d(F_{L}(l))}{dl}
=\frac{d_{c}-1}{2} [(P^{(u-1)}_{V^{nf}}(l)+P^{(u-1)}_{V^{f}}(l)\\+P^{(u-1)}_{V^{nf}}(-l)+P^{(u-1)}_{V^{f}}(-l))
 (\phi_{+}^{nf}(l)+\\ \phi_{+}^{f}(l)+\phi_{-}^{nf}(|l|)+\phi_{-}^{f}(|l|))^{d_{c}-2}].
\end{split}
\end{equation}
And for $l<0$,
\begin{equation}
\begin{split}
Q_{L}^{(u)}(l) =\frac{d_{c}-1}{2}[(P^{(u-1)}_{V^{nf}}(l)+P^{(u-1)}_{V^{f}}(l)-P^{(u-1)}_{V^{nf}}(-l)\\-P^{(u-1)}_{V^{f}}(-l))
(\phi_{+}^{nf}(l)+\phi_{+}^{f}(l)-\phi_{-}^{nf}(|l|)-\phi_{-}^{f}(|l|)^{d_{c}-2}].
\end{split}
\end{equation}
For irregular LDPC code, recall that the degree distribution pair
$\lambda$ and
$\rho$,
the check node procedure for DE is as follows.

For $l\geq0$,
\begin{equation}
\begin{split}
Q_{L}^{(u)}(l)=\sum^{d_{c}}_{d=1}\rho_{d} \frac{d-1}{2} [(P^{(u-1)}_{V^{nf}}(l)+P^{(u-1)}_{V^{f}}(l)+P^{(u-1)}_{V^{nf}}(-l)\\+P^{(u-1)}_{V^{f}}(-l)) 
(\phi_{+}^{nf}(l)+\phi_{+}^{f}(l)+\phi_{-}^{nf}(|l|)+\phi_{-}^{f}(|l|))^{d_{c}-2}].
\end{split}
\end{equation}
And for $l<0$,
\begin{equation}
\begin{split}
Q_{L}^{(u)}(l)=\sum^{d_{c}}_{d=1}\rho_{d} \frac{d-1}{2} [(P^{(u-1)}_{V^{nf}}(l)+P^{(u-1)}_{V^{f}}(l)-P^{(u-1)}_{V^{nf}}(-l)\\-P^{(u-1)}_{V^{f}}(-l))
(\phi_{+}^{nf}(l)+\phi_{+}^{f}(l)-\phi_{-}^{nf}(|l|)-\phi_{-}^{f}(|l|)^{d_{c}-2}].
\end{split}
\end{equation}
 The variable node procedure can be numerically computed using the Fast Fourier Transform(FFT).
\begin{equation}
P^{(u+1)}_{Z}(z)=FFT^{-1}\{\sum^{d_{v}}_{d=1}\lambda_{d}(FFT\{Q_{L}^{(u)}(l)\})^{d}\}.
\end{equation}
For the $(u+1)$-th iteration, the bit error probability $P_{e}$ can be calculated as follows:
\begin{equation}
P_{e}=\int_{-\infty}^{\infty}P^{(u+1)}_{Z}(z)dz.
\end{equation}
From (30) to (31), the DE algorithm for the proposed UEP scheme is obtained to evaluate
the performance of the UEP LDPC code. After the second iteration is processed, the flipped-bit and the non-flipped bit parts are merged. Hence, we apply an identical min-sum DE algorithm according to \cite{Jinghu} as follows.
 We have the check node procedure for DE is as follows:\\
 
 For $l\geq0$,
\begin{equation}
\begin{split}
Q_{L}^{(u)}(l)=\sum^{d_{c}}_{d=1}\rho_{d} \frac{d-1}{2}[(P^{(u-1)}_{V}(l)+P^{(u-1)}_{V}(-l))(\phi_{+}(l)\\+\phi_{-}(|l|))^{d_{c}-2}].
\end{split}
\end{equation}
And for $l<0$,
\begin{equation}
\begin{split}
Q_{L}^{(u)}(l)=\sum^{d_{c}}_{d=1}\rho_{d} \frac{d-1}{2}[(P^{(u-1)}_{V}(l)-P^{(u-1)}_{V}(-l))(\phi_{+}(l)\\-\phi_{-}(|l|))^{d_{c}-2}].
\end{split}
\end{equation}
The proposed DE algorithm is processed as follows.
\begin{description}
\item[Step 1:]\hspace{0.1cm}
Quantize the message $L_{d}(v_{i})$ at the input of the LDPC decoder following \cite{Chung} to obtain the error message of the pdf.
\item[Step 2:]\hspace{0.1cm}
Process the pdf using (30) to (31) for iteration $u$=0.
\item[Step 3:]\hspace{0.1cm}
Increase the value of $u$ by $1$ and process the pdf using (32), (34) and (35).
\item[Step 4:]\hspace{0.1cm}
If $u \geq u_{max}$ or $P_{e}<10^{-6}$ from (32), stop the algorithm, otherwise return to Step 3.
\end{description}
\subsection{Simulation Results}
\begin{figure}[hbt]
\centering
\captionsetup{font=scriptsize}
\includegraphics[width=0.5\textwidth]{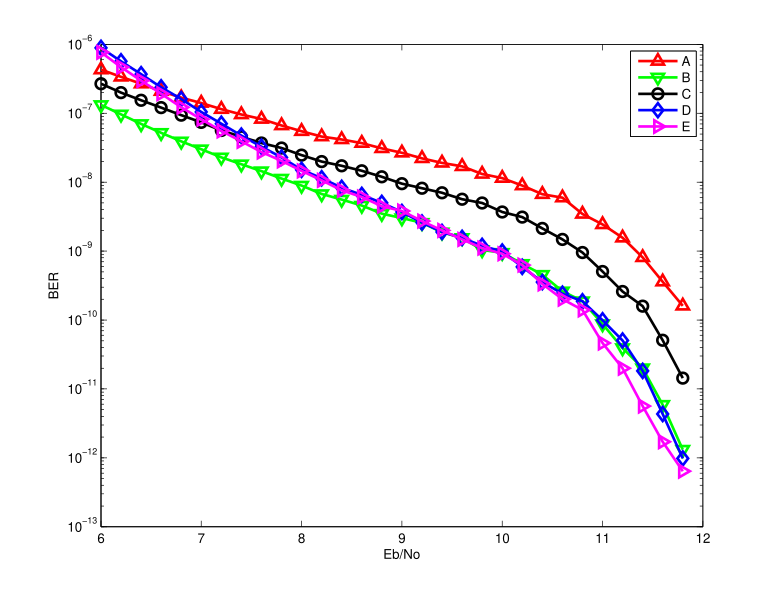}
\caption{Density evolution for 4-level $(0,3)$ RLL constraint over optical recording channel, where $\beta$ = 0.15, using (4608,3000) PEG-LDPC code rate 0.65. The proposed scheme type II is applied.
 (A) Code 5;
 (B) Code 2;
 (C) Code 1;
 (D) Code 3;
 (E) Code 4.
  }
\label{uep7}
\end{figure}

Since the occurrence of the RLL flipped bit is based on a random codeword, the decoding threshold is difficult to measure in conventional DE.
Hence, in order to present a more practical analysis, we set a number of DE iterations and then calculate the message error probability, which is substituted for searching theoretical
threshold. As illustrated in Fig.\,\ref{uep7}, $(4608,3000)$ LDPC code rate $0.65$
LDPC code with different code parameters is used and we set the maximum iteration of DE is equal to $15$. Curve (B) and Curve (C) present that $VND=[2,5]$ with $VND=[0.5,0.5]$
has better DE result than $VND=[2,3,5]$ with $\delta_{k}=[0.442,0.0874,0.4706]$, since Curve (B) has better node distribution where the proportion of optimum check node weight 10 is 0.9707 (Curve (B)) larger than 0.96782 (Curve (C)).  Moreover, Curve (B) is much better than Curve (D) and Curve (E) over medium SNR. Consequently, the insufficient error correcting capability causes an error floor over the flipped system. Thus, DE provides the prediction that Curve (D) and Curve (E) can perform slightly better than Curve (B) in high SNR. However, it is difficult to verify BER performance lower than $10^{-10}$. As a result, the DE result predicts that
$VND=[2,5]$ and $CND=[10]$ are the best node distributions for the RLL recording system and exhibit a satisfactory analysis for the code design over the symbol-dependent flipped system. This result is clear to explain the reason for the simulation result of Fig.\,\ref{uep5} and Fig.\,\ref{uep6}, since the minimum hamming distance predicted by DE for code 4 and code 5 is insufficient for the flipped system. Particularly for long code, Curve (E) in Fig.\,\ref{uep6} reveals good BER performance since the node distribution of Curve (E) is very close to the best node distribution $VND=[2,5]$ and $CND=[10]$, which is match to the DE result.

\begin{table*}[ht]\scriptsize
\centering
\caption{The node distribution of UEP LDPC code used in Fig.\,\ref{uep7}.}
\begin{tabular}[b]{|c|c|c|c|c|c|c|}
\hline
UEP Code&Code length&Code rate&VND&$\delta$&CND&$\gamma$ \\
\hline
Code 1&4608&0.65&[2,3,5]&[0.442,0.0874,0.4706]&[10,11]&[0.96782,0.03218] \\
\hline
Code 2&4608&0.65&[2,5]&[0.5,0.5]&[10,11]&[0.9707,0.0293]  \\
\hline
Code 3&4608&0.65&[2,6]&[0.5,0.5]&[11,12]&[0.95937,0.04063]  \\
\hline
Code 4&4608&0.65&[2,7]&[0.5,0.5]&[12,13,14]&[0.8691,0.02416,0.10674] \\
\hline
Code 5&4608&0.65&[2,4]&[0.5,0.5]&[8,9,10]&[0.3289,0.66692,0.00418]\\
\hline

\end{tabular}
\label{ueptable7}
\end{table*}

\subsection{EXIT Characteristics for Iterative Decoding Analysis}
Turbo equalization in \cite{Douillard} provides an efficient
decoding method at the reading side of typical recording systems,
and the EXIT (extrinsic information transfer) charts \cite{Brink}
are well-understood tools for observing the behavior of iterative
decoding. However, there are two differences in the EXIT
characteristics for flipped-bit recording systems compared with the
conventional approach: one is that we adapt the flipped operation
within the iterative decoding, and the other is that the iterative
decoding in \cite{Chen_Lin} utilizes the non-reset algorithm
\cite{Siegel}, which is different from the reset algorithm employed
in conventional EXIT chart analysis. 

We employ EXIT (extrinsic information transfer) characteristics
\cite{Brink} analysis to evaluate the information exchange between
the LDPC decoder and the MAP equalizer (detector).  Since there is
an LLR adjuster inserted between the MAP equalizer and the LDPC
decoder, the transfer functions of the MAP equalizer, and the LDPC
decoder is defined as follows.  The transfer
characteristic $T_Z$ for the MAP equalizer takes one-half of the LLR
adjuster into consideration.  Hence, the input mutual information is
$I_{Z,in} =I_{Z}(L_{a}(y_{i}),y_{i})$, and the output mutual
information is $I_{Z,out}$ = $I_{Z}(L_{a}(v_{i}),v_{i})$ =
$T_{Z}(I_{Z,in})$, where the mutual information is followed in \cite{Brink} as
 \begin{equation}
 I=\frac{1}{2}\sum_{x=-1,1}\int_{\infty}^{\infty}p(\xi|X=x)\times log\frac{2\cdot p(\xi|X=x)}{p(\xi|X=-1)+p(\xi|X=1)}d\xi.
 \end{equation}
 $p(\cdot )$ denote the LLR probability density function and $X$ is the transmitted massage.
 Note that no Gaussian assumption is imposed on the
 $p(\cdot )$ LLR probability density function.
 Since the MAP equalizer takes the channel
condition into consideration, $T_Z$ is dependent on $E_{b}/N_{0}$,
$\beta$, and $k$, which are the AWGN, jitter noise,
and RLL constraint parameters, respectively. The transfer characteristic $T_D$
for the LDPC decoder takes the other half of the LLR adjuster into
consideration. Hence, the input mutual information is  $I_{D,in} =
I_{D}(L_{a} (v_{i}),v_{i})$, and the output mutual information is
$I_{D,out}= I_{D}(L_{a}(y_{i}),y_{i})=T_{D}(I_{D,in})$.

\begin{figure}[hbt]
\centering
\captionsetup{font=scriptsize}
\includegraphics[width=0.5\textwidth]{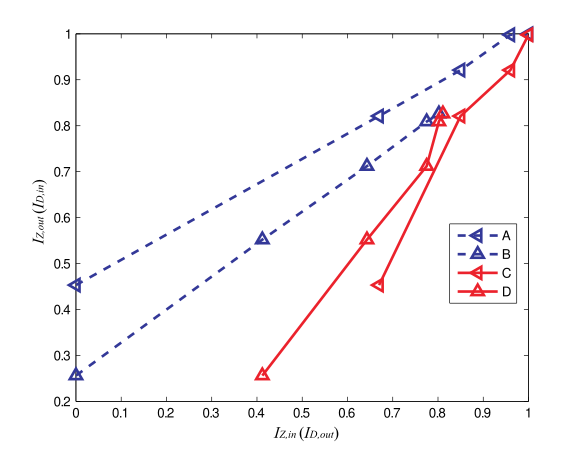}
\caption{EXIT characteristic for 4-level $(0,3)$ RLL constraint over optical recording channel, where $\beta$ = 0.15 and $E_b/N_0$ = 7.4 dB., using (4608,3000) PEG-LDPC code rate 0.65 with $U_{o}=5$ and $U_{i}=3$, using the non-reset decoder. The proposed scheme type II is applied.
 (A) $T_{Z}$ of Code 2;
 (B) $T_{Z}$ of Code 4;
 (C) $T_{D}$ of Code 2;
 (D) $T_{D}$ of Code 4.
  }
\label{uep11}
\end{figure}

In order to ensure that the error floor is eliminated and shows the convergence behavior, we are still interested in the convergent behavior for the proposed system.  In Fig.~\ref{uep11}, $L_{d}(v_{i})$ and $L_{D}(v_{i})$ are taken to calculate the pdf of input and output message. Hence, the input mutual information is
$I_{Z,in} =I_{Z}(L_{d}(v_{i}),v_{i})$, and the output mutual
information is $I_{Z,out}$ = $I_{Z}(L_{D}(v_{i}),v_{i})$ =
$T_{Z}(I_{Z,in})$. The transfer characteristic $T_D$
for the LDPC decoder takes the other half of the LLR adjuster into
consideration. Hence, the input mutual information is  $I_{D,in} =
I_{D}(L_{D} (v_{i}),v_{i})$, and the output mutual information is
$I_{D,out}= I_{D}(L_{d}(v_{i}),v_{i})=T_{D}(I_{D,in})$.
Code 2 and Code 4 are considered on the following EXIT characteristics.
In Fig.~\ref{uep11}, the jitter noise is considered for the practical channel.
The flipped system of Curves (A) and  (C) using Code 2 do converge to a mutual information point with a value
close to 1 at $E_b/N_0$ of 7.4 dB. However, the flipped system using Code 4 deteriorates to
reveal error floor regions and hold the mutual information far from 1.

\subsection{Optimization of the Node Distribution for the Proposed Scheme}
Finally, we present the optimization approach for the node distribution of the proposed scheme as follows. Linear programming for node distribution optimization can easily solve this problem.  However, the reasons that we do not apply linear programming to optimize the node distribution include (i) The node degree is unable to be considered the girth effect under code construction and leads to medium-length code performance degradation; (ii) The behavior of the occurrence of flipped bit is unable to depict under certain relation or model. Despite that the linear programming approach allows computing the optimum degree under Gaussian assumption, the flipped bit is still not ensured to be recovered; (iii) The linear programming is known as a local optimization process and needs to restart with different distribution until a good sequence is found. Then, differential evolution \cite{Storn} for global optimum distribution is applied to present the optimization approach as follows.
\begin{enumerate}
\item Initialization:\\
Set the smallest and largest number of category for $VND$ are $K^{g}_{v}=2$ and $K^{max}_{v}$ and generation index $g=0$. Since section IV has determined weight-7 as the maximum weight for variable node distribution, we apply the identical searching rule for the following steps that increase the maximum weight by $1$ until the code performance is getting worse, and then we stop increasing the maximum weight. Hence, the $K^{max}_{v}$ is determined by the maximum weight directly. Based on the condition of two levels of error protection, we randomly generate the variable node distribution $\delta^{g}_{l}$ with cardinality $L$ for $l=1,2,...,L$. The PEG construction is applied to construct an LDPC code and the check node distribution $\gamma$ is set. Then, process Monte Carlo simulation for the proposed flipping system to determine the error floor region.
Find the distribution denoted as $VND^{g}_{best}$ and $\delta^{g}_{l_{best}}$ has no error floor and the best BER performance for flipped system. In section III E, we find $VND^{0}_{best}=[2,5]$ and $\delta^{0}_{l_{best}}=[0.5,0.5]$
\item Mutation and test:\\
 Set $g=g+1$ and $K^{g+1}_{v}=K^{g}_{v}+1$. Generate the $VND^{g+1}$ from $I_{v}$ and $\delta^{g+1}_{l}=\delta^{g}_{l_{best}}+\alpha(\delta^{g}_{i}-\delta^{g}_{j}),i\neq j\neq l_{best},i,j\in {1,2,...,L}$ for $l=1,2,...,L$. Note that the $\alpha$ controls the amplification of the differential variation and the two levels of error protection must be held so that $I_{v_{weak}}=[2,3,4]$ and $I_{v_{strong}}=[4,5,6,7,even~larger]$ with distribution equally 0.5. Also, find the maximum weight to determine $K^{max}_{v}$ in this category.
\item Compare and update:\\
For $l=1,2,...,L$, if the BER performance is better $\delta^{g+1}_{l_{best}}=\delta^{g+1}_{l}$ and $VND^{g+1}_{best}=VND^{g+1}$, otherwise $\delta^{g+1}_{l_{best}}=\delta^{g}_{l_{best}}$ and $VND^{g+1}_{best}=VND^{g}$
\item Stopping test and output:\\ 
 If $K^{g}_{v}$ reach the $K^{max}_{v}$, we stop searching. Otherwise, go to Step 2.
\end{enumerate}
It is interesting to note that the control variable of DE includes $\alpha$ and $L$. According to \cite{Storn}, the author suggests that the initial value of $\alpha=0.5$ is usually a good choice. If the population converges prematurely, then $\alpha$
and/or $L$ should be increased. Additionally, the value of $L$ is determined by experience. We set $L=50$.
We organize all possible trial node distribution simulating over the theoretical optical recording channel and the error floor status is our great concern for the recording system. Additionally, every trial node distribution is simulated using $U_{o}=5$ and $U_{i}=3$ for code length 4608 and $U_{o}=5$ and $U_{i}=5$ for code length 11520 under the practical condition for the limited number of iterations. There are three viewpoints to reduce the searching space organized as follows: (i) In section III, we observe that the check node weight-10 constructs the best check code for the proposed scheme under code rate 0.65 discussed in section IV, the degree weight of searching space is limited; (ii) The concentration theorem is proven to ensure all codes in an
ensemble have the same behavior; (iii) The error floor for the proposed scheme is caused by the insufficient error correcting capability in the strong part which the node proportion is less than 0.5. Furthermore, in order to improve the system performance, increasing the correction capability in the strong part is effective for proposed scheme type II, since the strong part possesses the labeling with less number of $NN$. Particularly, long code length possesses more flipping errors and the error floor status is identical to the medium code length. As a result, the BER performance of $VND^{0}_{best}=[2,5]$ and $\delta^{0}_{l_{best}}=[0.5,0.5]$ is improved by means of maintaining the check node weight-10 and slightly adding some proportion of weight-7 and more weight in the strong part. This consideration is confirmed by the search for differential evolution. Otherwise, performance enhancement is not presented by adding more weight to the weak part. We point out some good $VND$ performing better than  $VND^{0}_{best}=[2,5]$ which $VND=[2,5,7]~\delta=[0.5,0.499,0.001]$, $VND=[2,5,7,8]~\delta=[0.5,0.498,0.001,0.001]$, $VND=[2,5,7,8,9]~\delta=[0.5,0.497,0.001,0.001,0.001]$. An even more higher proportion of large weight results in the check node being larger than the best check code of weight 10. Finally, the best node distribution using differential evolution for code length 4608 and 10240 is $VND^{g}_{best}=[2,5,7,8]$ and $\delta^{g}_{l_{best}}=[0.5,0.498,0.001,0.001]$.

\begin{table*}[ht]\scriptsize
\centering
\caption{The trial node distribution of UEP LDPC code for differential evolution optimization.}
\begin{tabular}[b]{|c|c|c|c|c|c|c|}
\hline
Code length&Code rate&VND&$\delta$&CND&$\gamma$&Error floor \\
\hline
4608&0.65&[2,5]&[$<$0.5,$>$0.5]&[10,11]&N.A.&No  \\
\hline
4608&0.65&[2,5]&[0.5,0.5]&[10,11]&[0.9707,0.0293]&No  \\
\hline
4608&0.65&[2,6]&[0.5,0.5]&[11,12]&[0.95937,0.04063]&Yes  \\
\hline
4608&0.65&[2,6]&[$\tau$,$<$0.5 or $>$0.5]&N.A&N.A&Yes \\
\hline
4608&0.65&[2,7]&[0.5,0.5]&[12,13,14]&[0.8691,0.02416,0.10674]&Yes \\
\hline
4608&0.65&[2,7]&[$\tau$,$>$0.5 or $<$0.5]&N.A&N.A&Yes \\
\hline
4608&0.65&[2,4]&[0.5,0.5]&[8,9,10]&[0.3289,0.66692,0.00418]&Yes\\
\hline
4608&0.65&[3,4]&[0.5,0.5]&[10,11]&[0.96794,0.032]&Yes\\
\hline
4608&0.65&[3,5]&[0.5,0.5]&[11,12]&[0.5169,0.4831]&No\\
\hline
4608&0.65&[2,3,5]&[0.442,0.0874,0.4706]&[10,11]&[0.96782,0.03218]&Yes \\
\hline
4608&0.65&[2,3,5]&[$\tau$,$<$0.5,$>$0.5]&[10,11]&N.A&Yes \\
\hline
4608&0.65&[2,3,5]&[$\tau$,$<$0.5,$>$0.5]&[11,12]&N.A&Yes \\
\hline
4608&0.65&[2,3,5]&[$\tau$,$\tau$-0.5,0.5]&[10,11]&N.A&No \\
\hline
4608&0.65&[2,3,6]&[$\tau$,$\tau$-0.5,0.5]&[11,12]&N.A&Yes \\
\hline
4608&0.65&[2,3,7]&[$\tau$,$\tau$-0.5,0.5]&[12,13]&N.A&Yes\\
\hline
4608&0.65&[2,3,4]&[$\tau$,$\tau$-0.5,0.5]&[9,10]&N.A&Yes \\
\hline
4608&0.65&[2,4,5]&[$\tau$,$\tau$-0.5,0.5]&[11,12]&N.A&No \\
\hline
4608&0.65&[2,4,5]&[0.5,$\tau$,0.5-$\tau$]&[9,10]&N.A&No \\
\hline
4608&0.65&[2,4,5]&[$\tau$,0.5,0.5-$\tau$]&[9,10]&N.A&Yes \\
\hline
4608&0.65&[2,4,6]&[0.5,$\tau$,0.5-$\tau$]&[9,10]&N.A&Yes \\
\hline
4608&0.65&[2,4,7]&[0.5,$\tau$,0.5-$\tau$]&[9,10]&N.A&Yes \\
\hline
4608&0.65&[2,5,6]&[0.5,$\tau$,0.5-$\tau$]&[9,10]&N.A&No \\
\hline
4608&0.65&[2,5,7]&[0.5,$\tau$,0.5-$\tau$]&[9,10]&N.A&No \\
\hline
4608&0.65&[2,3,4,5]&[$\tau$,0.5-$\tau$,@,0.5-@]&[9,10]&N.A&Yes \\
\hline
4608&0.65&[2,3,4,5]&[0.5-$\tau$-@,$\tau$,@,0.5]&[9,10]&N.A&No \\
\hline
4608&0.65&[2,3,5,6]&[$\tau$,0.5-$\tau$,@,0.5-@]&[9,10]&N.A&No \\
\hline
4608&0.65&[2,3,5,7]&[$\tau$,0.5-$\tau$,@,0.5-@]&[9,10]&N.A&No \\
\hline
4608&0.65&[2,3,6,7]&[$\tau$,0.5-$\tau$,@,0.5-@]&[9,10]&N.A&Yes \\
\hline
4608&0.65&[2,4,5,6]&[$\tau$,0.5-$\tau$,@,0.5-@]&[10,11]&N.A&Yes \\
\hline
4608&0.65&[2,4,5,7]&[$\tau$,0.5-$\tau$,@,0.5-@]&[10,11]&N.A&Yes \\
\hline
4608&0.65&[2,5,6,7]&[0.5,$\tau$,@,0.5-@-$\tau$]&[10,11]&N.A&No \\
\hline
\end{tabular}
\label{ueptable2}
\end{table*}

\begin{table*}[ht]\scriptsize
\centering
\caption{The trial node distribution of UEP LDPC code for differential evolution optimization.}
\begin{tabular}[b]{|c|c|c|c|c|c|c|}
\hline
Code length&Code rate&VND&$\delta$&CND&$\gamma$&Error floor \\
\hline
4608&0.65&[2,5,6,8]&[0.5,$\tau$,@,0.5-@-$\tau$]&[10,11]&N.A&No \\
\hline
4608&0.65&[2,5,7,8]&[0.5,$\tau$,@,0.5-@-$\tau$]&[10,11]&N.A&No \\
\hline
4608&0.65&[2,5,7,9]&[0.5,$\tau$,@,0.5-@-$\tau$]&[10,11]&N.A&No \\
\hline
4608&0.65&[2,5,7,10]&[0.5,$\tau$,@,0.5-@-$\tau$]&[10,11]&N.A&No \\
\hline
4608&0.65&[2,3,4,5,6]&[$\tau$,0.5-$\tau$,@,0.5-@-$\kappa$,$\kappa$]&[9,10]&N.A&Yes \\
\hline
4608&0.65&[2,3,4,5,6]&[$\tau$,0.5-$\tau$-@,@,0.5-$\kappa$,$\kappa$]&[9,10]&N.A&No \\
\hline
4608&0.65&[2,4,5,7,8]&[$\tau$,0.5-$\tau$,@,0.5-@-$\kappa$,$\kappa$]&[10,11]&N.A&Yes \\
\hline
4608&0.65&[2,4,5,7,9]&[$\tau$,0.5-$\tau$,@,0.5-@-$\kappa$,$\kappa$]&[10,11]&N.A&Yes \\
\hline
4608&0.65&[2,4,5,7,10]&[$\tau$,0.5-$\tau$,@,0.5-@-$\kappa$,$\kappa$]&[10,11]&N.A&Yes \\
\hline
4608&0.65&[2,4,5,8,9]&[$\tau$,0.5-$\tau$,@,0.5-@-$\kappa$,$\kappa$]&[10,11]&N.A&Yes \\
\hline
4608&0.65&[2,4,5,8,10]&[$\tau$,0.5-$\tau$,@,0.5-@-$\kappa$,$\kappa$]&[10,11]&N.A&Yes \\
\hline
4608&0.65&[2,5,7,8,9]&[0.5,$\tau$,@,$\kappa$,0.5-$\kappa$-$\tau$-@]&[10,11]&N.A&No \\
\hline
4608&0.65&[2,5,7,8,10]&[0.5,$\tau$,@,$\kappa$,0.5-$\kappa$-$\tau$-@]&[10,11]&N.A&No \\
\hline
4608&0.65&[2,5,7,8,9,10]&[0.5,$\tau$,@,$\kappa$,$\vartheta$,0.5-$\kappa$-$\tau$-@-$\vartheta$]&[10,11]&N.A&No \\
\hline
\end{tabular}
\label{ueptable2}
\end{table*}

\begin{table*}[ht]\scriptsize
\centering
\caption{The trial node distribution of UEP LDPC code for differential evolution optimization.}
\begin{tabular}[b]{|c|c|c|c|c|c|c|}
\hline
Code length&Code rate&VND&$\delta$&CND&$\gamma$&Error floor \\
\hline
11520&0.65&[2,5]&[0.5,0.5]&[9,10]&[0.00074,0.99926]&No\\
\hline
11520&0.65&[2,6]&[0.5,0.5]&[10,11]&[0.85615,0.14385]&Yes\\
\hline
11520&0.65&[2,7]&[0.5,0.5]&[10,11]&[0.713,0.287]&Yes\\
\hline
11520&0.65&[2,6]&[$\tau$,$<$0.5 or $>$0.5]&N.A&N.A&Yes \\
\hline
11520&0.65&[2,7]&[$\tau$,$<$0.5 or $>$0.5]&N.A&N.A&Yes \\
\hline
11520&0.65&[2,3,5]&[$\tau$,$<$0.5,$<$0.5]&[10,11]&N.A&Yes \\
\hline
11520&0.65&[2,3,5]&[$\tau$,$<$0.5,$>$0.5]&[11,12]&N.A&Yes \\
\hline
11520&0.65&[2,3,5]&[$\tau$,$\tau$-0.5,0.5]&[10,11]&N.A&No \\
\hline
11520&0.65&[2,3,6]&[$\tau$,$\tau$-0.5,0.5]&[11,12]&N.A&Yes \\
\hline
11520&0.65&[2,3,7]&[$\tau$,$\tau$-0.5,0.5]&[12,13]&N.A&Yes\\
\hline
11520&0.65&[2,3,4]&[$\tau$,$\tau$-0.5,0.5]&[9,10]&N.A&Yes \\
\hline
11520&0.65&[2,4,5]&[$\tau$,$\tau$-0.5,0.5]&[11,12]&N.A&No \\
\hline
11520&0.65&[2,4,5]&[0.5,$\tau$,0.5-$\tau$]&[9,10]&N.A&No \\
\hline
11520&0.65&[2,4,5]&[$\tau$,0.5,0.5-$\tau$]&[9,10]&N.A&Yes \\
\hline
11520&0.65&[2,4,6]&[0.5,$\tau$,0.5-$\tau$]&[9,10]&N.A&Yes \\
\hline
11520&0.65&[2,4,7]&[0.5,$\tau$,0.5-$\tau$]&[9,10]&N.A&Yes \\
\hline
11520&0.65&[2,5,6]&[0.5,$\tau$,0.5-$\tau$]&[9,10]&N.A&No \\
\hline
11520&0.65&[2,5,7]&[0.5,$\tau$,0.5-$\tau$]&[9,10]&N.A&No \\
\hline
11520&0.65&[2,3,4,5]&[$\tau$,0.5-$\tau$,@,0.5-@]&[9,10]&N.A&Yes \\
\hline
11520&0.65&[2,3,4,5]&[0.5-$\tau$-@,$\tau$,@,0.5]&[9,10]&N.A&No \\
\hline
11520&0.65&[2,3,5,6]&[$\tau$,0.5-$\tau$,@,0.5-@]&[9,10]&N.A&No \\
\hline
11520&0.65&[2,3,5,7]&[$\tau$,0.5-$\tau$,@,0.5-@]&[9,10]&N.A&No \\
\hline
11520&0.65&[2,3,6,7]&[$\tau$,0.5-$\tau$,@,0.5-@]&[9,10]&N.A&Yes \\
\hline
11520&0.65&[2,4,5,6]&[$\tau$,0.5-$\tau$,@,0.5-@]&[10,11]&N.A&No \\
\hline
11520&0.65&[2,4,5,7]&[$\tau$,0.5-$\tau$,@,0.5-@]&[10,11]&N.A&No \\
\hline
11520&0.65&[2,4,5,8]&[$\tau$,0.5-$\tau$,@,0.5-@]&[10,11]&N.A&No \\
\hline
11520&0.65&[2,4,5,9]&[$\tau$,0.5-$\tau$,@,0.5-@]&[10,11]&N.A&No \\
\hline
11520&0.65&[2,4,5,10]&[$\tau$,0.5-$\tau$,@,0.5-@]&[10,11]&N.A&No \\
\hline
11520&0.65&[2,5,6,7]&[0.5,$\tau$,@,0.5-@-$\tau$]&[10,11]&N.A&No \\
\hline
\end{tabular}
\label{ueptable2}
\end{table*}

\begin{table*}[ht]\scriptsize
\centering
\caption{The trial node distribution of UEP LDPC code for differential evolution optimization.}
\begin{tabular}[b]{|c|c|c|c|c|c|c|}
\hline
Code length&Code rate&VND&$\delta$&CND&$\gamma$&Error floor \\
\hline
11520&0.65&[2,5,6,8]&[0.5,$\tau$,@,0.5-@-$\tau$]&[10,11]&N.A&No \\
\hline
11520&0.65&[2,5,7,8]&[0.5,$\tau$,@,0.5-@-$\tau$]&[10,11]&N.A&No \\
\hline
11520&0.65&[2,5,7,9]&[0.5,$\tau$,@,0.5-@-$\tau$]&[10,11]&N.A&No \\
\hline
11520&0.65&[2,3,4,5,6]&[$\tau$,0.5-$\tau$,@,0.5-@-$\kappa$,$\kappa$]&[10,11]&N.A&Yes \\
\hline
11520&0.65&[2,3,4,5,7]&[$\tau$,0.5-$\tau$-@,@,0.5-$\kappa$,$\kappa$]&[10,11]&N.A&Yes \\
\hline
11520&0.65&[2,3,4,5,8]&[$\tau$,0.5-$\tau$-@,@,0.5-$\kappa$,$\kappa$]&[10,11]&N.A&Yes \\
\hline
11520&0.65&[2,3,4,5,9]&[$\tau$,0.5-$\tau$-@,@,0.5-$\kappa$,$\kappa$]&[10,11]&N.A&Yes \\
\hline
11520&0.65&[2,4,5,6,7]&[0.5,$\tau$,@,$\kappa$,0.5-@-$\tau$-$\kappa$]&[10,11]&N.A&Yes \\
\hline
11520&0.65&[2,4,5,7,8]&[0.5,$\tau$,@,$\kappa$,0.5-@-$\tau$-$\kappa$]&[10,11]&N.A&Yes \\
\hline
11520&0.65&[2,4,5,8,9]&[0.5,$\tau$,@,$\kappa$,0.5-@-$\tau$-$\kappa$]&[10,11]&N.A&Yes \\
\hline
11520&0.65&[2,5,7,8,9]&[0.5,$\tau$,@,$\kappa$,0.5-$\kappa$-$\tau$-@]&[10,11]&N.A&Yes \\
\hline
11520&0.65&[2,5,7,8,10]&[0.5,$\tau$,@,$\kappa$,0.5-$\kappa$-$\tau$-@]&[10,11]&N.A&Yes \\
\hline
11520&0.65&[2,5,7,8,9,10]&[0.5,$\tau$,@,$\kappa$,$\vartheta$,0.5-$\kappa$-$\tau$-@-$\vartheta$]&[10,11]&N.A&Yes \\
\hline
\end{tabular}
\label{ueptable2}
\end{table*}

\begin{figure}[hbt]
\captionsetup{font=scriptsize}
\includegraphics[width=0.5\textwidth]{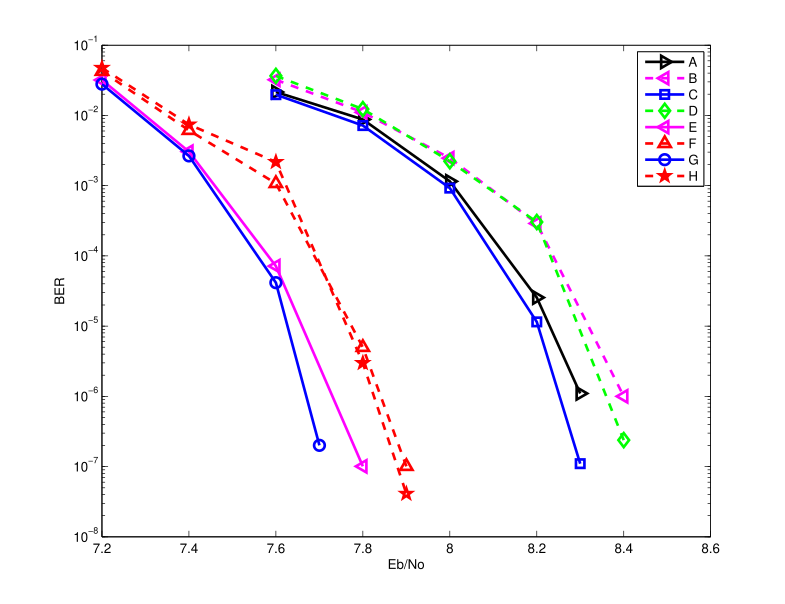}
\caption{BER results for 4-level $(0,3)$ RLL constraint over PR channel $\overline{h}$=$(1,2,2,1)$ using PEG-LDPC code with rate 0.65. The proposed scheme type II is applied. For Curve (A) to Curve (D), $U_{o}=5$ and $U_{i}=3$ is applied and while for Curve (E) to Curve (H), $U_{o}=5$ and $U_{i}=5$ is applied.
 (A) Non-flipped system using Code 2;
 (B) Flipped system using Code 2;
 (C) Non-flipped system using Code 13;
 (D) Flipped system  using Code 13;
 (E) Non-flipped system using Code 10;
 (F) Flipped system using Code 10;
 (G) Non-flipped system using Code 14;
 (H) Flipped system using Code 14;
  }
\label{uep12}
\end{figure}

\begin{figure}[hbt]
\captionsetup{font=scriptsize}
\includegraphics[width=0.5\textwidth]{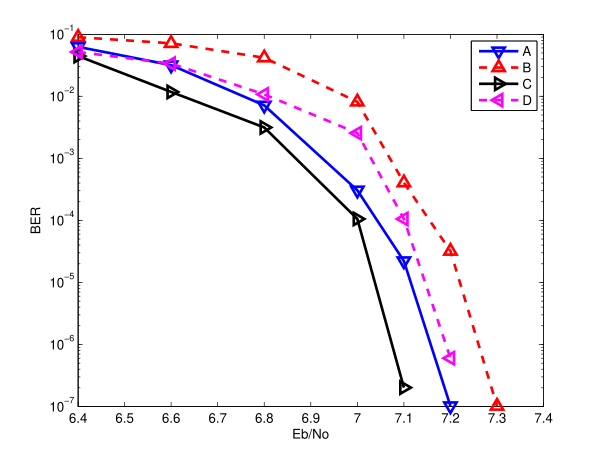}
\caption{BER results for 4-level $(0,3)$ RLL constraint over optical recording channel using PEG-LDPC code with rate 0.65, where $\beta$ = 0.15 and $U_{o}=5$ and $U_{i}=3$. The proposed scheme type II is applied.
 (A) Non-flipped system using Code 2;
 (B) Flipped system using Code 2;
 (C) Non-flipped system using Code 13;
 (D) Flipped system using Code 13.
  }
\label{uep14}
\end{figure}

Finally, we illustrate the optimum distribution for the proposed flipped system
in Fig.~\ref{uep12}. For code length 4608, Code 13 of the variable weight 7 and 8 result in slightly worse performance in medium SNRs and better performance in high SNR as shown in the flipped system of Curve (D). The Code 13 reveals better high SNR performance than Code 2. For code length 11520, Curve (H) reveals identical result. Furthermore, jitter noise is also considered in Fig.~\ref{uep14}. Code 13 has better performance result comparing to Code 2. Consequently, we remind the density evolution predicts that the variable weight higher than 5 has better BER performance in high SNR region, which induct to approach the optimum distribution by differential evolution.

\begin{table*}[ht]\scriptsize
\centering
\caption{The node distribution of UEP LDPC code used in Fig.\,\ref{uep12}.}
\begin{tabular}[b]{|c|c|c|c|c|c|c|}
\hline
UEP Code&Code length&Code rate&VND&$\delta$&CND&$\gamma$ \\
\hline
Code 2&4608&0.65&[2,5]&[0.5,0.5]&[10,11]&[0.9707,0.0293]  \\
\hline
Code 10&11520&0.65&[2,5]&[0.5,0.5]&[9,10]&[0.00074,0.99926]\\
\hline
Code 13&4608&0.65&[2,5,7,8]&[0.5,0.498,0.001,0.001]&[10,11]&[0.954,0.0456]  \\
\hline
Code 14&11520&0.65&[2,5,7,8]&[0.5,0.498,0.001,0.001]&[10,11]&[0.99945,0.00055]\\
\hline
\end{tabular}
\label{ueptablexx}
\end{table*}

In order to optimize LDPC node distribution, density evolution (DE) allows us to achieve the best decoding threshold. However, the RLL flipped bit is generated from a random codeword, and difficult to depict the behavior.
Then, we apply the DE algorithm for performance evaluation instead of searching the decoding threshold. For simplifying the implementation of numerical analysis, min-sum-based DE is applied. The original min-sum-based DE is derived into flipped-bit and non-flipped bit parts of message pdfs. We present a DE algorithm for the symbol-dependent channel with an RLL flipped bit. Although the min-sum algorithm has performance loss compared with the sum-product algorithm, the proposed min-sum density evolution is still functional to identify the LDPC code performance between different node distributions.
According to section IV at the expense of Monte Carlo simulation, we observe that $VND=[2,x]$-based is an interesting ensemble distribution to investigate and recommend the LDPC code with $VND=[2,5]$ with $\delta=[0.5,0.5]$ is the best for the proposed system.
Based on the DE result, the LDPC code with $VND=[2,6]$ and $[2,7]$ has more powerful correcting on repetition codes and leads to more sharp slope, however, the check code is weaker than the code with $VND=[2,5]$. Hence, the check code weight-10 is better than the others over the proposed flipped system which results in performing better on the DE result between 10$^{-6}$ to 10$^{-10}$ . Consequently, the insufficient error correction capability of the codes with $VND=[2,6]$ and $[2,7]$ leads to perform error floor. Additionally, this prediction of the proposed DE result is accurate not only for long LDPC code but also for medium code length code.  EXIT characteristic is also applied to ensure that the error floor is eliminated. The mutual information is converged to a value close to 1. Finally, differential evolution is applied to prove the best distribution for the proposed flipped system. We conclude the condition of recovering the flipped bit is that the Euclidean separation is necessary to increase for a certain level so that the flipped system has sufficient correction capability to eliminate the error floor.

\section{Conclusion}
In this paper, we first present the recording system with RLL-constraint
and the flipped-bit detection method for deliberating the flipping system. We discuss the conventional multilevel coded modulation for unequal protection and the system performance of signal labeling using Ungerboeck partitioning and block partitioning. The study indicates that reducing the number of nearest neighbors ($NN$) at every partition level allows for strengthening the UEP capability of different levels. The design strategy for maximizing intraset distance at each level is not enough for constructing modulation codes with UEP property. In addition to the larger difference of Euclidean separation between each level, block partitioning remains a large decrease in effective error coefficients in order to prevent error propagation. Then, we observe that irregular LDPC code is the convenient approach to UEP error correcting capability. The PEG construction applied in this thesis is a modern computer-based construction for searching large girth LDPC code. We present a UEP LDPC-coded scheme using deliberate flipping over 4-level RLL $(0,k)$ constraint. The regular interleaver allows the flipped-bit to appear limited on the specific segment of the codeword. The proposed UEP scheme exploits unequal protection property efficiently on the flipped bit part to alleviate the interference of flipped bits. Based on the design strategy of multilevel UEP modulation, a well-designed UEP-coded system allows to recovery of the flipped bit and mitigates the high error floor. Simulation results presented different LDPC code rates and the performance of the proposed scheme approaching nearly to the non-flipped system.
In order to obtain the performance evaluation, density evolution(DE) for the proposed system is derived and provides a numerical analysis for the code design instead of searching decoding threshold. We provide a practical tool to analyze the node distribution over a symbol-dependent system. The DE simulation result indicates an interesting prediction for the possible node distribution and EXIT chart reveals the assurance of recovering a flipped bit. Finally, we prove the optimum distribution using the differential evolution for the proposed recording system. Finally, we conclude these two topics of this thesis. If the $k$ of the RLL $(d,k)$ constraint is not very small, the flipped-bit detection approach for the flipped system has better BER performance than the UEP LDPC coding scheme. However, the RLL constraint becomes more strict which causes unsolvable erroneous detection and leads to an error floor. Thus at the expense of the weak correction part of codes, the flipped system of the UEP coding scheme using optimum distribution is more robust to recover the strict-constraint flipped bit by the strong correction part.
\bibliography{main}
\bibliographystyle{unsrt}

\balance

\end{document}